\documentclass[traditabstract]{aa}

\usepackage{graphicx}
\usepackage{txfonts}
\usepackage{color}
\usepackage[switch,modulo]{lineno}
\usepackage{rotating}
\usepackage{xspace}
\usepackage{float}
\usepackage{url}

\usepackage{gensymb}
\usepackage{threeparttable}
\graphicspath{{./}}

\newcommand{\g}{$\gamma$\xspace}
\newcommand{\wco}{$W_{\rm{CO}}$\xspace}
\newcommand{\wcounit}{K km s$^{-1}$\xspace}

\newcommand{\hd}{H$_2$\xspace}
\newcommand{\cosat}{$\rm{CO_{\rm sat}}$\xspace}
\newcommand{\hi}{{\sc Hi}\xspace}
\newcommand{\hii}{{\sc Hii}\xspace}
\newcommand{\nh}{$N_{\rm{H}}$\xspace}
\newcommand{\nhi}{$N_{\rm{HI}}$\xspace}

\newcommand{\nhd}{$N_{\rm{H}_2}$\xspace}

\newcommand{\nhgam}{$N_{\rm{H}\,\gamma}$\xspace}

\newcommand{\opa}{$\tau_{353}/N_{\rm{H}}$\xspace}
\newcommand{\opavghi}{$\overline{\tau_{353}/N}_{\rm{H}}^{\rm{HI}}$\xspace}
\newcommand{\opavgco}{$\overline{\tau_{353}/N}_{\rm{H}}^{\rm{CO}}$\xspace}
\newcommand{\opaunit}{$10^{-27}$ cm$^2$ H$^{-1}$\xspace}
\newcommand{\av}{$A_{\rm{V}}$\xspace}
\newcommand{\ebv}{$E$(B$-$V)\xspace}

\newcommand{\ebvnh}{$E$(B$-$V)/$N_{\rm H}$\xspace}
\newcommand{\ebvnhgam}{$\widetilde{E}({\rm B-V})$/$N_{\rm{H}\,\gamma}$\xspace}
\newcommand{\ebvnhunit}{$10^{-22}$~mag~cm$^2$\xspace}
\newcommand{\ebvtau}{$E$(B$-$V)/$\tau_{353}$\xspace}
\newcommand{\avtau}{$A_{\rm{V}}/\tau_{353}$\xspace}

\newcommand{\anh}{$A_{\rm{V}}/N_{\rm{H}}$\xspace}

\newcommand{\xco}{$X_{\rm{CO}}$\xspace}

\newcommand{\xcounit}{$10^{20}$ cm$^{-2}$ K$^{-1}$ km$^{-1}$ s\xspace}
\newcommand{\taunu}{$\tau_{353}$\xspace}
\newcommand{\taunuavg}{$\overline{\tau_{353}}$\xspace}

\newcommand{\anaE}{``$\gamma{+}E$(B$-$V)''\xspace}
\newcommand{\anaT}{``$\gamma{+}\tau_{353}$''\xspace}

\begin{document}

   \title{Cosmic-rays, gas, and dust in nearby anti-centre clouds : \\ III -- Dust extinction, emission, and grain properties }

\author{
Q.~Remy$^{(1,2)}$ \and 
I.~A.~Grenier$^{(1)}$ \and 
D.J.~Marshall$^{(1)}$ \and 
J.~M.~Casandjian$^{(1)}$ 
}
\authorrunning{LAT collaboration}

\institute{
\inst{1}~Laboratoire AIM, CEA-IRFU/CNRS/Universit\'e Paris Diderot, Service d'Astrophysique, CEA Saclay, F-91191 Gif sur Yvette, France\\
\inst{2}~Laboratoire Univers et Particules de Montpellier, Universit\'e Montpellier, CNRS/IN2P3, F-34095 Montpellier, France\\ 
\email{quentin.remy@umontpellier.fr} \\
\email{isabelle.grenier@cea.fr} \\
}

   \date{Received 1 July 2017 / Accepted 13 March 2018}
   
\abstract
{}
{
{We have explored the capabilities of dust extinction and \g rays to probe the properties of the interstellar medium in the nearby anti-centre region.
In particular, we aim at quantifying the variations of the dust properties per gas nucleon across the different gas phases and different clouds. The comparison of dust extinction and emission properties with other physical quantities of large grains (emission spectral index $\beta$, dust colour temperature $T_{\rm dust}$,  total-to-selective extinction factor $R_{\rm V}$) helps the theoretical modelling of grains as they evolve from diffuse to dense cloud environments.}
} 
{We have jointly modelled the \g-ray intensity, recorded between 0.4 and 100 GeV with the \textit{Fermi} Large Area Telescope ({LAT}), and the stellar reddening, \ebv, inferred from \textit{Pan-STARRS} and \textit{2MASS} photometry, as a combination of \hi-bright, CO-bright, and ionised gas components. The complementary information from dust reddening and \g rays is used to reveal the gas not seen, or poorly traced, by \hi, free-free, and $^{12}$CO emissions, namely (i) the opaque \hi and diffuse \hd present in the Dark Neutral Medium (DNM) at the atomic-molecular transition, and (ii) the dense \hd to be added where $^{12}$CO lines saturate (\cosat). We compare the total gas column densities, \nh, derived from the \g rays and  stellar reddening with those inferred from a similar analysis \citep{2017A&A...601A..78R} of \g rays and of the optical depth of the thermal dust emission, \taunu, at 353 GHz. We can therefore compare environmental variations in specific dust reddening, \ebvnh, and in dust emission opacity (dust optical depth per gas nucleon), \opa.}
{The gas column densities obtained when combining \g rays with either dust reddening or dust emission compare reasonably well in the atomic and DNM gas phases and over most of the CO-bright phase, but we find localized differences in the dense media (\cosat component) due to differences in the two dust tracers. Over the whole anti-centre region, we find an average \ebvnh ratio of $(2.02\pm0.48)\times$\ebvnhunit, with maximum local variations of about $\pm30\%$ at variance with the two to six fold coincident increase seen in emission opacity as the gas column density increases. We show how the specific reddening and opacity vary with the colour temperature and spectral index of the thermal emission of the large grains.
Additionally, we find a better agreement between the \mbox{\xco $=N($\hd)/\wco} conversion factors derived with dust reddening or with \g rays than with those inferred from dust emission, especially toward clouds with large \taunu optical depths. The comparison confirms that the high \xco values found with dust emission are biased by the significant rise in emission opacity inside molecular clouds.}
{In the diffuse medium, we find only small variations in specific reddening, \ebvnh, compatible with the dispersion in the $R_{\rm V}$ factor reported by other studies. This implies a rather uniform dust-to-gas mass ratio in the diffuse parts of the anti-centre clouds.  The small amplitude of the \ebvnh variations with increasing \nh column density confirms that the large opacity \opa rise seen toward dense CO clouds is primarily due to changes in dust emissivity. The environmental changes are qualitatively compatible with model predictions based on mantle accretion on the grains and the formation of grain aggregates. }

\titlerunning{gas \& dust in nearby anti-centre clouds III}

\maketitle

\section{Introduction}

Large dust grains in the interstellar medium (ISM) absorb starlight from the UV to near infrared domain causing appreciable extinction and reddening in stellar observations \citep{2003ApJ...598.1017D}. These grains in thermal equilibrium with the ambient radiation field re-emit in the infrared to millimetre wavelengths. If the dust grains are well mixed with the interstellar gas, the reddening they cause and their thermal emission both provide indirect tracers of the gas column density.

The reddening caused by dust grains is generally estimated from the photometry of a set of stars in a given direction. A family of methods is based on the comparison of the observed stellar colours in various photometric bands with reference colours corresponding to zero extinction \citep[NICE, NICER, NICEST algorithms:][]{1994ApJ...429..694L,2001A&A...377.1023L,2009A&A...493..735L}. Such colour-excess methods have been applied to \textit{2MASS} photometry to derive two-dimensional maps of the dust reddening over the whole sky \citep{2009MNRAS.395.1640R,2016A&A...585A..38J}. \cite{2006A&A...453..635M} developed a method to estimate simultaneously the dust reddening and distance by comparing the observed colour from \textit{2MASS} data with the intrinsic colours and distances of stars provided by the Besancon Galaxy model \citep{2003A&A...409..523R}. More recently, \cite{2015ApJ...810...25G} have used the combined information of \mbox{\mbox{\textit{Pan-STARRS} 1}} and \textit{2MASS} surveys to derive a three-dimensional reddening map in a fully probabilistic framework described in \cite{2014ApJ...783..114G}. 
  
The thermal emission of the large grains has been recorded at 100 and 240 $\mu$m with \textit{IRAS} and \textit{DIRBE} and has been used to estimate dust optical depths over the whole sky \citep{1998ApJ...500..525S}. This map has been scaled to \ebv colour excesses using a set of quasar reddening measurements, but its structure reflects the distribution of optical depths. In a more recent study \citep{2014A&A...571A..11P}, the emission measured with \textit{Planck} and \textit{IRAS} between 353 and 3000 GHz (100 to 850 $\mu$m) has been used to model the spectral energy distributions of the thermal emission and to derive an all-sky map of the dust optical depth in emission at 353 GHz, \taunu. The latter relates to dust column density and grain emissivity. Assuming uniform dust properties (size distribution, chemical composition, and grain structure), the dust optical depth can be converted into extinction or reddening.

In addition to the information provided by dust tracers, the \g rays with energies above few hundred MeV that are produced by the interaction of cosmic rays (CRs) with gas nucleons provide a measure of the total gas column density under the assumption of a uniform CR flux through a given cloud \citep{1978A&A....70..367C,1982A&A...107..390L,1988A&A...207....1S}. As shown in \cite{2017A&A...601A..78R}, the interstellar \g-ray spectra measured in the nearby clouds of the Galactic anti-centre region studied here are consistent with a uniform CR flux across the region and with a uniform penetration of the cosmic rays with energies above a few GeV through the different gas phases of the clouds, from the atomic envelopes to the molecular cores. Therefore we can use \g rays as a robust tracer of the total gas column density, \nh, in this region.

Several studies have noted an increase in dust opacity (optical depth per gas nucleon, \opa) with increasing gas column density from the atomic to molecular phases. In the Taurus clouds, the opacities increase in the dense CO filaments by a factor ranging from ${\sim}2$ \citep{2009ApJ...701.1450F,2011A&A...536A..25P,2013A&A...559A.133Y} to $3.4^{+0.3}_{-0.7}$ \citep{2003A&A...398..551S} above the diffuse-ISM value. In a study of the Galactic anti-centre region including the Taurus clouds \citep{2017A&A...601A..78R}, we have found opacity enhancements reaching a factor of six toward the CO clouds, even though we provided a measure of the additional \hd gas that is not linearly traced by CO intensity, \wco, in the directions where the $^{12}$CO line emission saturates (\cosat). We also found that the opacity starts to increase in the Dark Neutral Medium (DNM) at the \hi--\hd transition, therefore at lower gas densities than the few thousand per cm$^3$ sampled in the bulk of the CO-bright volume. The grain opacity found in the DNM compares well in the anti-centre clouds \citep{2017A&A...601A..78R} and in the Chamaeleon complex \citep{2015A&A...582A..31A}. 

The opacity increase could be caused by an evolution of the grain properties changing their mass emission coefficient, as suggested by recent theoretical works \citep{2015A&A...579A..15K}, or by spatial variations in the dust-to-gas mass ratio. The latter possibility also impacts the relation between dust reddening and gas column density.

This paper is a follow up study of our previous work \citep[paper I: ][]{2017A&A...601A..78R}, which focused on the analysis of interstellar \g rays and dust thermal emission in the nearby anti-centre clouds. 
We aim here to study variations in the distribution of \ebvnh ratios, across the different gas phases and in different types of clouds, using \g rays to trace the total gas column density. In order to directly compare environmental changes in \ebvnh and in \opa, we use the same region, sampling, analysis method, and gas data as before. 

The various ISM tracers used in this analysis and the expressions that relate \ebvnh and \opa to dust properties are introduced in Sec. \ref{sec:data}. The joint analysis of dust and \g-ray tracers is detailed in \cite{2017A&A...601A..78R} and summarized in Sec. \ref{sec:mod} and \ref{sec:iter}. The \ebvnh and \opa ratios are derived from fits of dust and \g-ray models to the data, as described in Sec. \ref{sec:method}. The results of the joint \anaE analysis are given in Sec. \ref{sec:res} and compared to the results of the former \anaT analysis. In Sec. \ref{sec:ccl}, we compare our results to the expectations of recent theoretical models of dust grain evolution \citep{2013A&A...558A..62J,2015A&A...579A..15K} and discuss how the variations in \opa and \ebvnh relate to changes in the other physical quantities characterizing large grains. The consequences for the capability of these dust tracers to probe the gas column density are discussed in Sec. \ref{sec:ccl}. 

Additionally, we compare the reddening maps that are currently available and comment on the choice made for this analysis in Appendix \ref{sec:AVcomp}. We present the best-fit coeffcients of the dust and \g-ray models in Appendix \ref{sec:annex}.

\begin{figure*}
  \centering               
  \includegraphics[scale=0.55]{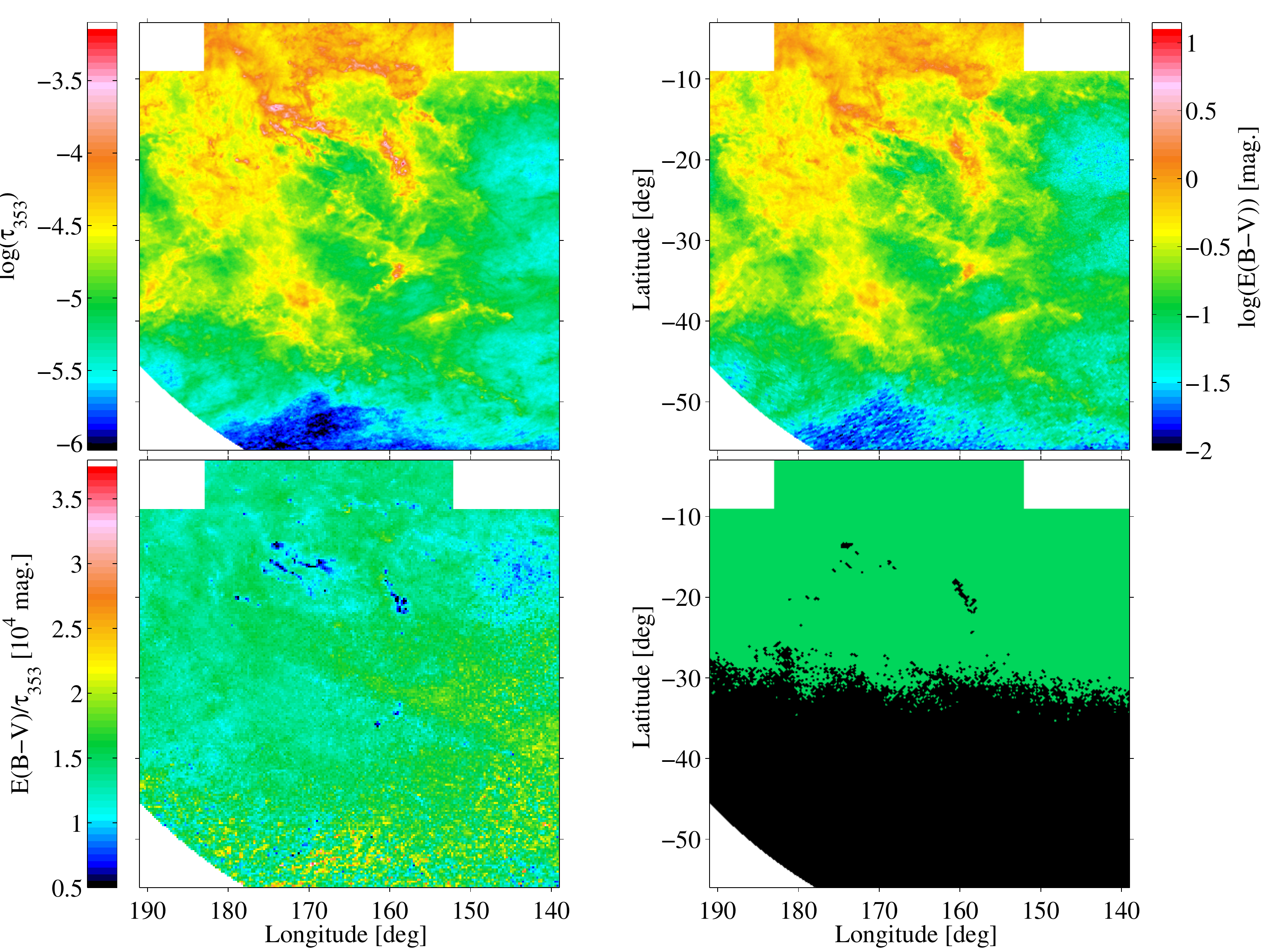}
\caption{Maps of the dust optical depth at 353 GHz (top left), of the stellar reddening by dust (top right), and of their ratio (bottom left). The map on the lower right shows in green the pixels with an angular resolution of 0\fdg11 in \ebv, which have been retained for mask2, and in black the pixels with a lower resolution (see Sec. \ref{sec:masks}).}  
  \label{fig:Dustmaps}
\end{figure*}

\begin{figure*}
  \centering               
  \includegraphics[scale=0.6]{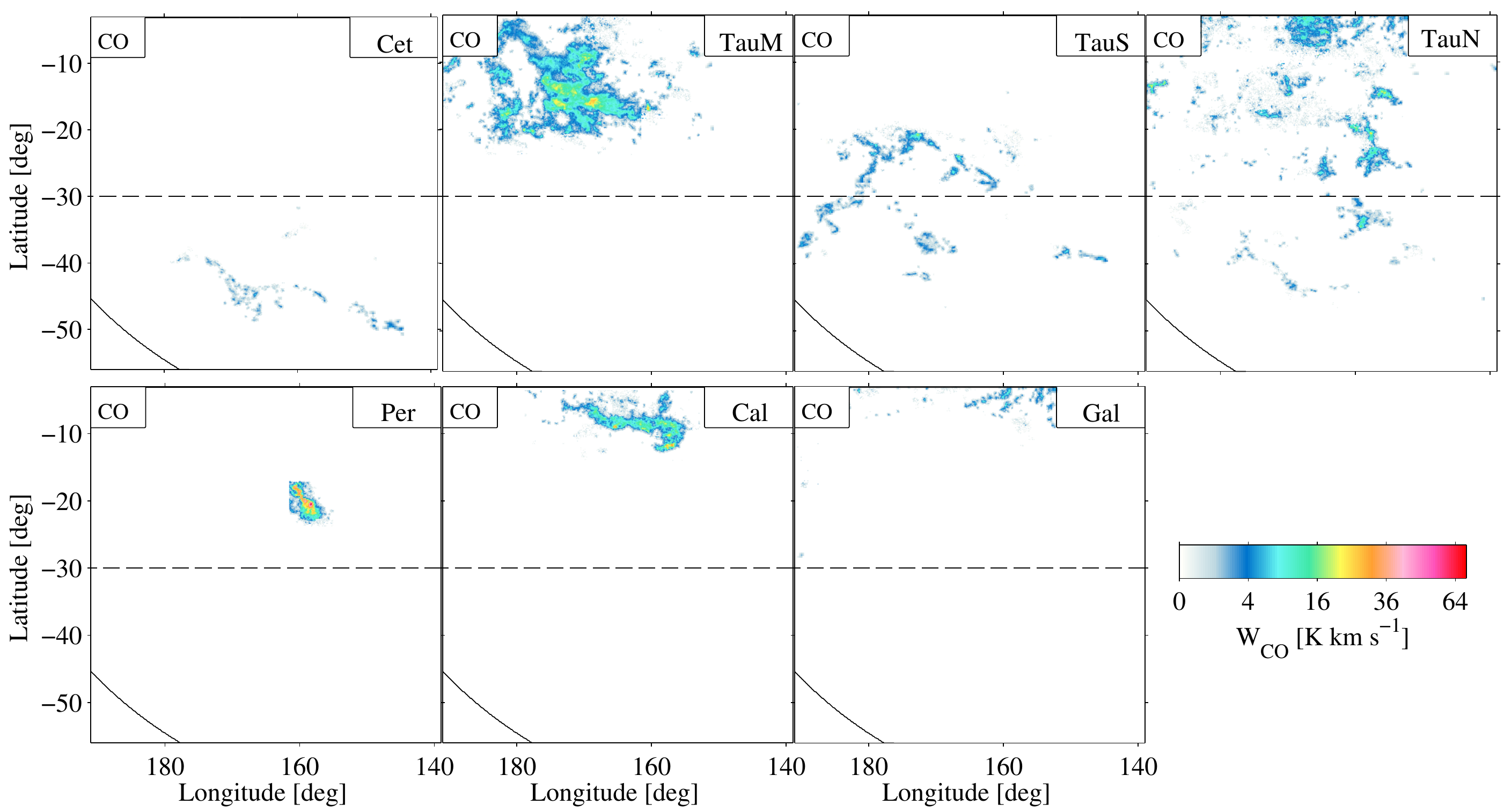}
  \caption{CO intensity maps of the Cetus (Cet), Main Taurus (TauM), South Taurus (TauS), North Taurus (TauN), Perseus (Per), and California (Cal) clouds, and of the Galactic disc background (Gal). The solid curve and dashed line show the low-latitude borders of two analysis regions, respectively mask 0 and mask1. }
  \label{fig:COmaps}
\end{figure*}

\section{Data} \label{sec:data}

For our analyses, we have selected the same anti-centre region as that studied in \cite{2017A&A...601A..78R}.  It extends from 139\degr\, to 191\degr\, in Galactic longitude and from $-56\degr$ to $-3\degr$ in latitude. Regions with large gas column densities from the Galactic disc in the background and regions with a coarser angular resolution in the \hi map have been masked. The analysis region encompasses a variety of clouds, from diffuse and mostly atomic entities to dense molecular clouds forming stars at different rates, and spanning distances between 140 and 410 pc from the Sun \cite[see Table 1 of paper II: ][]{2017arXiv171105506R}.

\subsection{Dust emission and extinction} \label{sec:def}

The dust optical depth is defined as the product of the dust emission cross section per unit mass, $\kappa_\nu$ in cm$^{2}$ g$^{-1}$, and the dust mass column density, $M_{\rm dust}$. The emission cross section can be expressed as $\kappa_\nu$=$\kappa_0\left( \frac{\nu}{\nu_0} \right)^\beta $ with $\kappa_0$ the value at the reference frequency $\nu_0$ and $\beta$ the spectral index \citep{1983QJRAS..24..267H,2011A&A...525A.103C,2014A&A...571A..11P}. The dust mass column density is expressed as a function of the hydrogen column density as  $M_{\rm dust}=R_{\rm DG}\mu_{\rm H} N_{\rm H} $ with $\mu_{\rm H} = 2.27 \times 10^{-27}$ kg the mean gas mass per hydrogen atom, and  $R_{\rm DG}$ the dust-to-gas mass ratio.
So the dust emission optical depth can be expressed as: 
\begin{equation}\label{eq:taunu}
\tau_\nu= \kappa_\nu M_{\rm dust} =\kappa_{\nu_0}\left( \frac{\nu}{\nu_0}\right)^\beta R_{\rm DG}\mu_{\rm H}N_{\rm H} 
\end{equation}
and then we can define the dust emission opacity or hereafter in short, opacity \citep{2014A&A...571A..11P}, as: 
\begin{equation}\label{eq:taunh}
\frac{\tau_\nu}{N_{\rm H}}=\kappa_{\nu_0}\left( \frac{\nu}{\nu_0}\right)^\beta R_{\rm DG}\mu_{\rm H}
\end{equation}

The dust extinction, $A_{\lambda}$, at a wavelength $\lambda$ is proportional to the extinction optical depth along a line of sight, thus to the integral of the dust volume density, $n_{\rm dust}$, and of the extinction cross section $\sigma^{\rm ext}_{\lambda}$ \citep{1998ppim.book.....S}:
\begin{equation}
A_{\lambda}=1.086 \,\tau^{\rm ext}_{\lambda}=1.086 \int n_{\rm dust}\sigma^{\rm ext}_{\lambda} \rm ds
\end{equation}
Assuming the grains can be described by spheres of effective radius $a$ and mean mass density $\rho_{\rm gr}$, we have $\sigma^{\rm ext}_{\lambda}=Q_\lambda^{\rm ext}\pi a^2$ with $Q_\lambda^{\rm ext}$ the efficiency factor for extinction. So the extinction optical depth can be expressed as a function of the hydrogen column density as:
\begin{equation}
\tau^{\rm ext}_{\lambda}= Q_\lambda^{\rm ext}\pi a^2 N_{\rm dust}
= Q_\lambda^{\rm ext} \frac{3\mu_{\rm H}}{4\rho_{\rm gr} a} R_{\rm DG}N_{\rm H}
\end{equation}
Substituting this relation in the expression for the visual extinction, $A_{\rm V}$, and using the definition of the total-to-selective extinction factor $R_{\rm V}=\frac{A_{\rm V}}{E(\rm{B}-\rm{V})}$ yields to an expression of the specific reddening: 

\begin{equation}\label{eq:ebvnh}
\frac{E(\rm{B}-\rm{V})}{N_{\rm H}}=1.086 \frac{3Q_{\rm V}^{\rm ext}}{4 \rho_{\rm gr} a R_{\rm V}} R_{\rm DG}\mu_{\rm H}
\end{equation}

A relation between the optical depth and the reddening can be obtained by combining equations \eqref{eq:taunu} and \eqref{eq:ebvnh}: 
\begin{equation}\label{eq:ebvtau}
\frac{\tau_\nu}{E(\rm{B}-\rm{V})}= 1.228 \,\frac{\rho_{\rm gr} a }{Q_{\rm V}^{\rm ext}} \,R_{\rm V} \,\kappa_{\nu_0}\left( \frac{\nu}{\nu_0}\right)^\beta 
\end{equation}

Provided we can reliably trace the total gas column densities in all gas forms, variations in \opa and in \ebvnh can probe intrinsic variations in dust properties and in the dust-to-gas mass ratio. The knowledge of the gas column density and cloud type can inform us on environmental effects on dust evolution in the different gas phases.

We have used the dust optical depth at 353 GHz, \taunu, from \cite{2014A&A...571A..11P}. It has been derived by fitting the spectral energy distribution of the dust emission between 353 and 3000 GHz (100 to 850 $\mu$m) on a 5\arcmin scale, as measured with \textit{Planck} and \textit{IRAS}.
 
The stellar reddening map used in the following comes from the three-dimensional \ebv map of \cite{2015ApJ...810...25G}, integrated up to its maximal distance. This map has been derived using \mbox{\textit{Pan-STARRS} 1} and \textit{2MASS} stellar data.
We have tested other reddening maps to quantify the differences induced by the use of other stellar data and different methods. As the reddening map of \cite{2015ApJ...810...25G} presents a better correlation with both the \taunu optical depth and the gas column densities inferred in \g rays, we use this map to study the relation between \ebv and \nh (see the discussion on the choice of extinction map in Appendix \ref{sec:AVcomp}).

The distributions of optical depth \taunu, reddening \ebv, and of their \ebvtau ratio are displayed for the region of interest in Fig. \ref{fig:Dustmaps}. It shows that regions of intermediate \ebv ($\sim$0.3--1 mag) have a ratio close to the value of 1.49$\times 10^4$ mag found by \cite{2014A&A...571A..11P}. Yet, the ratio drops by a factor up to three at large extinctions in the direction of compact molecular clumps and filaments (further discussed in Sec. \ref{sec:ccl}).

\subsection{\g-ray data} \label{sec:gam}
We have used the same \g-ray data as in \cite{2017A&A...601A..78R}, namely six years of Pass 8 photon data, recorded between 0.4 and 100 GeV by the \textit{Fermi} Large Area Telescope \citep[LAT, ][]{2009ApJ...697.1071A}. We have used the associated instrument response functions (P8R2\_CLEAN\_V6). 
In order to reduce the contamination of residual CR tracks in the photon data and by Earth atmospheric \g rays, we have applied tight rejection criteria \citep[CLEAN class selection, photon arrival directions within < 100$^\circ$ of the Earth zenith and in time intervals when the LAT rocking angle was inferior to 52$^\circ$; for details, see][]{2012ApJS..199...31N}.
To account for the spill-over of emission produced outside the analysis region, but reconstructed inside it, we have modelled point sources and interstellar contributions in a region 4$^\circ$ wider than the analysis region.

The positions and spectra of the \g-ray sources in the field are provided by the \textit{Fermi}-LAT Third Source Catalog \citep{2015ApJS..218...23A}. The observed $\gamma$-ray emission also includes a contribution from the large-scale Galactic inverse Compton (IC) emission emanating from the interactions of CR electrons with the interstellar radiation field (ISRF). The GALPROP\footnote{\url{http://galprop.stanford.edu}} parameter file 54-LRYusifovXCO4z6R30-Ts150-mag2 has been  used to generate an energy-dependent template of the Galactic IC emission across the analysis region \citep{2012ApJ...750....3A}.
Details on the isotropic spectrum for the extragalactic and residual charged-particle backgrounds are given in \cite{2013arXiv1303.3514A} and \cite{2015ApJS..218...23A}.

\subsection{Gaseous components}\label{sec:gasdata}

We have used the same data as in \cite{2017A&A...601A..78R} to describe the gaseous components. We will briefly summarize their properties in the following, and more details can be found in the aforementioned paper.

The bright atomic gas is traced by the 21 cm line emission of hydrogen. 74\% of the analysis region is covered by the GALFA-\hi survey at 4\arcmin-resolution \citep{2011ApJS..194...20P}. Elsewhere we have used the EBHIS survey at 10.8\arcmin-resolution \citep{2016A&A...585A..41W}. Both surveys were re-sampled into the 0\fdg125-spaced Cartesian grid used for the analyses. We have verified the tight correlation in column densities between the two surveys where both are available in our analysis region \citep{2017A&A...601A..78R}.

In order to trace the bright molecular gas, we have used $^{12}$CO $J{=}1{\rightarrow}0$ observations at 115 GHz from the moment-masked CfA CO survey of the Galactic plane \citep{2001ApJ...547..792D,2004ASPC..317...66D}. Most of the survey is based on a 0\fdg125-spaced Cartesian grid except for the high-latitude clouds, at $b\lesssim-50\degr$, which have been interpolated from 0\fdg25 to 0\fdg125.

\hi and CO lines have been used to kinematically separate cloud complexes along the lines of sight and to separate the nearby clouds from the Galactic backgrounds. We have used the narrow-band GALFA data-cubes with their original velocity resolution of 0.18 km/s in the local standard of rest. The EBHIS frequency sampling is coarser, with a velocity resolution of \mbox{1.44 km s$^{-1}$}, but still adequate to separate clouds in velocity. Details of the pseudo-Voigt line decompositions and selection of longitude, latitude, and velocity boundaries of the different complexes are given in \cite{2017A&A...601A..78R}. The separation resulted in six different complexes for both \hi and CO components: Cetus, Main Taurus, South Taurus, North Taurus,  Perseus, and California. They are shown in CO intensity, \wco, in Fig. \ref{fig:COmaps} and in \nhi column density in Fig. 1 of \cite{2017A&A...601A..78R}. 

Ionised gas in the California nebula (around $l=160$\fdg1 and $b =-12$\fdg4), in the G159.6-18.5 \hii region, and along the Eridanus loop is visible in H$\alpha$ emission \citep{2003ApJS..146..407F} and in free-free emission at mm wavelengths \citep{2013ApJS..208...20B}. We have used the \textit{Planck} LFI data at 70 GHz \citep[release 2.01 data,][]{2015arXiv150201582P} with an angular resolution of 14\arcmin\, to measure the free-free emission in the analysis region. To do so, we have successively removed from the LFI data the contributions from the Cosmic Microwave Background, from dust emission, and from point sources \citep[see][ for details]{2017A&A...601A..78R}.

\section{Method and Analyses} \label{sec:method}

\subsection{Dust and \g-ray models} \label{sec:mod}

The large dust grains responsible for the stellar reddening at near IR and optical wavelengths are taken to be well mixed with the interstellar gas, so we have modelled the spatial distributions of the \ebv reddening as a linear combination of gas column densities in the various phases and from the different cloud complexes. The model is expressed similarly to  Eq. 6 of \cite{2017A&A...601A..78R}, substituting \taunu for \ebv:

\begin{eqnarray}
E(\rm{B-V})(l,b)&=& \sum_{i=1}^7 y_{\rm{HI},i} N_{\rm{HI},i}(l,b) + \sum_{i=1}^7 y_{\rm{CO},i} W_{\rm{CO},i}(l,b)  \notag\\
 &+&y_{\rm{ff}}I_{\rm{ff}}(l,b)  + y_{\rm{DNM}} N_{\rm H}^{\rm{DNM}}(l,b)  \notag\\
 &+&y_{\rm{COsat}} N_{\rm H}^{\rm{COsat}}(l,b) + y_{\rm{iso}},
\label{eq:modDust}
\end{eqnarray} 
where $N_{\rm{HI},i}(l,b)$, $W_{\rm{CO},i}(l,b)$, and $I_{\rm{ff}}(l,b)$ respectively denote the \nhi, \wco, and free-free maps of the clouds in the seven complexes. 
$N_{\rm H}^{\rm{DNM}}(l,b)$ and $N_{\rm H}^{\rm{COsat}}(l,b) $ stand for the column densities of the dark gas in the DNM and \cosat components. They are deduced from the coupled analyses of the \g-ray and dust data (see Sec. \ref{sec:iter}). The $y$ model parameters have been estimated using a $\chi^{2}$ minimization.

At the energies relevant for the LAT observations, the particle diffusion lengths in the interstellar medium exceed the cloud dimensions and there is no spectral indication of variations in CR density inside the clouds \citep{2017A&A...601A..78R}. The emissivity spectrum of the gas follows the average one obtained in the local ISM, $q_{\rm LIS}$, given by \cite{2015ApJ...806..240C}. So the interstellar \g radiation can be modelled, to first order, as a linear combination of gas column-densities in the various phases and different regions seen along the lines of sight. The model also includes a contribution from the large-scale Galactic inverse-Compton emission, $I_{\rm IC}(l,b,E)$, an isotropic intensity to account for the charged-particle and extragalactic backgrounds, $I_{\rm iso}(l,b,E)$, point sources, $S_j(E)$ and contamination of sources present in the peripheral region outside the analysis region merged in a single map, $S_{ext}(l,b,E)$.
The \g-ray intensity $I(l,b,E)$, expressed in cm$^{-2}$ s$^{-1}$ sr$^{-1}$ MeV$^{-1}$, can be modelled as : 
\begin{eqnarray}
      I(l,b,E) &=& q_{\rm LIS}(E) \,\times \, \Biggr[\, \sum_{i=1}^7 q_{\rm{HI},i}(E) \, N_{\rm{HI},i}(l,b) \nonumber\\
      &+&\, \sum_{i=1}^7 q_{\rm{CO},i}(E) \, W_{\rm{CO},i}(l,b) + q_{\rm{ff}}(E)I_{\rm{ff}}(l,b) \nonumber \\ 
      &+&\, q_{\rm{DNM}}(E) \,E(\rm{B}-\rm{V})^{\rm{DNM}}(l,b)\nonumber \\
       &+&\, q_{\rm{COsat}}(E)\, E(\rm{B}-\rm{V})^{\rm{COsat}}(l,b)\,\,\Biggr]\nonumber \\
      &+&\, q_{\rm{IC}}(E) \, I_{\rm IC}(l,b,E) + q_{\rm iso}(E) \, I_{\rm iso}(E) \nonumber\\
       &+&\, \sum_j q_{S_j}(E) \, S_j(E) \, \delta(l-l_j,b-b_j) \nonumber\\
        &+&q_{S{\rm ext}}(E) \, S_{\rm ext}(l,b,E), 
\label{eq:modGam}
\end{eqnarray} 
with $E(\rm{B}-\rm{V})^{\rm{DNM}}$ and $E(\rm{B}-\rm{V})^{\rm{COsat}}$ the reddening maps in the DNM and \cosat phases, extracted from the coupled dust and \g-ray analyses (see Sec. \ref{sec:iter}). 

To ensure photon statistics robust enough to follow details in the spatial distributions of the different interstellar components, we have analysed the data in five broad and independent energy bands, bounded by 10$^{2.6} $, 10$^{2.8} $, 10$^{3.2} $, 10$^{3.6} $, and 10$^{5} $ MeV and in the total energy band. We have left free normalisations (the $q$ factors) to account for possible deviations in CR density and spectrum for the interstellar components and to account for possible changes in source spectra since we have used a longer time interval of LAT data than for the 3FGL catalogue. Cloud-to-cloud variations in CR flux are monitored by the $q_{\rm{HI},i}$ scale factors and will affect the normalizations equally in all energy bands, whereas a change in CR penetration in a specific cloud will be revealed as an energy-dependent correction. For each cloud, the average \g-ray emissivity spectrum per H atom in the atomic phase is estimated from the product of the $q_{\rm LIS}(E)$ spectrum and the best-fit $q_{\rm{HI},i}(E)$ normalizations. This emissivity can be used to estimate the gas mass present in the other DNM, CO, and \cosat parts of the cloud if one assumes a uniform CR flux across the whole structure.

\begin{figure}
  \centering
  \includegraphics[width=\hsize]{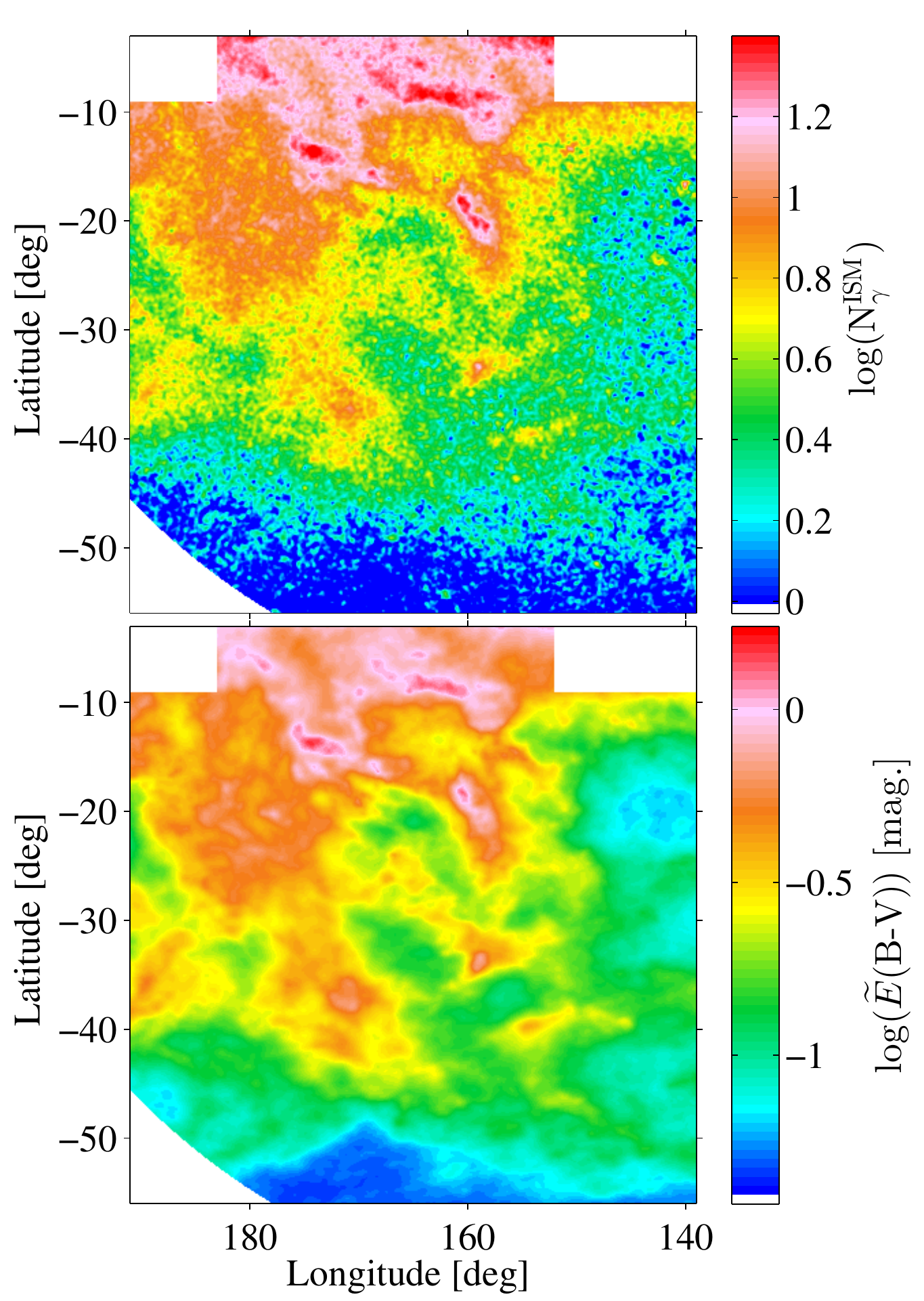}
  \caption{\textit{Top:} \g-ray counts of gaseous origin recorded in the 0.4--100 GeV energy band in a 0\fdg125 pixel grid. \g-ray emissions other than due to cosmic-ray interactions in the gas have been subtracted. The map has been smoothed with a Gaussian kernel of 0\fdg14 dispersion for display. \textit{Bottom:} stellar reddening by dust displayed at the \textit{Fermi}-LAT angular resolution for comparison.}  
  \label{fig:gamDust}
\end{figure}

Figure \ref{fig:gamDust} shows the distribution of \g rays of interstellar origin in the region, together with the stellar reddening displayed at the angular resolution of the \textit{Fermi} LAT. The \g rays of interstellar origin are obtained  through \g-ray data analysis after subtraction of the inverse Compton, isotropic, and discrete source contributions (see Eq. \ref{eq:modGam}). Fig. \ref{fig:gamDust} illustrates striking similarities between the spatial distributions of the stellar reddening by dust and of the gas traced by CR interactions. The same likeness had been found between the dust optical depth \taunu and the \g rays \citep[see Fig. 3 of][]{2017A&A...601A..78R}. The spatial correlation supports the use of dust tracers to complement the \hi and CO data in modelling the \g rays of gaseous origin. Yet, as for the optical depth \taunu, small differences between the two maps in Fig. \ref{fig:gamDust} readily point to local variations in specific reddening \ebvnh.

\subsection{Mapping DNM and \cosat components} \label{sec:iter}



Both the stellar reddening and interstellar \g-ray emission trace the total gas to first order. They are modelled as linear combinations of several gaseous components including bright (\textit{i.e.}, detected in radio or millimeter emission lines) neutral gas in the atomic and molecular phases (\hi, CO emissions), ionised gas (free-free emission), and dark gas. 
The analysis couples the dust and \g-ray models to constrain the structure of the additional gas not seen in \hi, CO and free-free emissions, namely the gas in the DNM at the \hi-\hd transition and the \cosat \hd gas missed by the saturation of the $^{12}$CO lines at large optical thickness in the dense cores of molecular clouds.
Both components show up as positive residuals over the expectations from the \hi-bright, CO-bright, and free-free-bright gas components. To extract
them, we have built maps of the positive residuals between the data (dust or \g-rays) and the best-fit contributions from the \nhi, \wco, free-free, and ancillary (other than gas) components. We have kept only the positive residuals above the noise. We have separated the DNM and \cosat components in directions outside and inside of the 7 \wcounit contour in \wco intensity.

The DNM and \cosat templates of additional gas used in the \g-ray model are inferred from the dust model and conversely. We have cycled between the dust and \g-ray fits iteratively until the quality (log-likelihood) of the fit to the \g-ray data saturated. The spatial correlation between the \g-ray and dust excesses ensured by this iteration is essential to extract genuine gas quantities and discard local variations in CR density or in dust properties. 
The hydrogen column densities in the DNM and \cosat gas phases have been estimated by using the \g-ray emissivity per hydrogen atom in the \hi phase and assuming that the DNM and \cosat phases are pervaded by the same average CR flux as in the atomic envelope. This assumption is validated by the uniformity of the \g-ray emissivity spectrum (\textit{i.e.}, CR spectrum) found in the different gas phases, including the dense CO and \cosat phases. By comparing the results of the \g-ray analyses using \taunu and \ebv as dust input, we can check for any small bias in the determination of hydrogen column densities in \g rays, which would be indirectly induced by the dust data used to infer the DNM and \cosat distributions.

\subsection{Choice of analysis masks}\label{sec:masks}


The \ebv map has been constructed with a resolution of 0\fdg11 or 0\fdg23 depending on the number of stars available in each pixel. Directions with a resolution below 0\fdg11 are shown in black in Fig. \ref{fig:Dustmaps}. They correspond to low latitudes far from the densely populated regions near the Galactic plane  ($b<-30\degr$) and to directions toward highly obscured molecular clouds.
In the very diffuse environments below -30\degr \,in latitude where \ebv<0.3 mag, the fluctuations seen in the \ebvtau ratio result from uncertainties in the zero level in \taunu, fluctuations from the cosmic infrared background in the \taunu map \citep{2014A&A...571A..11P}, and from an increased level of noise in \ebv at low stellar surface densities \citep{2015ApJ...810...25G}.
In order to compare dust emission and reddening on the same scale and to test the impact of the change in angular resolution across the \ebv map, we have performed the analysis with three different masks:
\begin{itemize}
\item mask0, including the full region displayed in Fig. \ref{fig:gamDust} (as for the \taunu and \g-ray analysis of \cite{2017A&A...601A..78R}),
\item mask1, excluding latitudes below $-30$\degr\, to keep regions with high signal-to-noise in \ebv (dotted cut in Fig. \ref{fig:COmaps}),
\item mask2, excluding all the pixels with an angular resolution in \ebv worse than 0\fdg11 (displayed in black in Fig. \ref{fig:Dustmaps}).
\end{itemize}

The California, Perseus, and Taurus molecular clouds are fully kept in mask1, but Cetus is masked out in both mask1 and mask2. Those masks discard 30$\%$ and 18$\%$ of the pixels with CO emission in the South and North Taurus clouds, respectively. Fifteen percent of the pixels toward the Perseus cloud are masked out with mask2. Since the \mbox{CO-to-\hd} conversion factor (\xco= \nhd/\wco) tends to decrease from the edge to the core of molecular clouds \citep{2006MNRAS.371.1865B,2016MNRAS.455.3763B,2017A&A...601A..78R} masking out parts of a cloud changes the mean \xco value of the cloud compared to the result obtained for the complete cloud (mask0). This effect is discussed in Sec. \ref{sec:resxco}.

 \begin{figure}[!ht]
 \vspace{0.6cm}
  \centering   
  \includegraphics[width=\hsize]{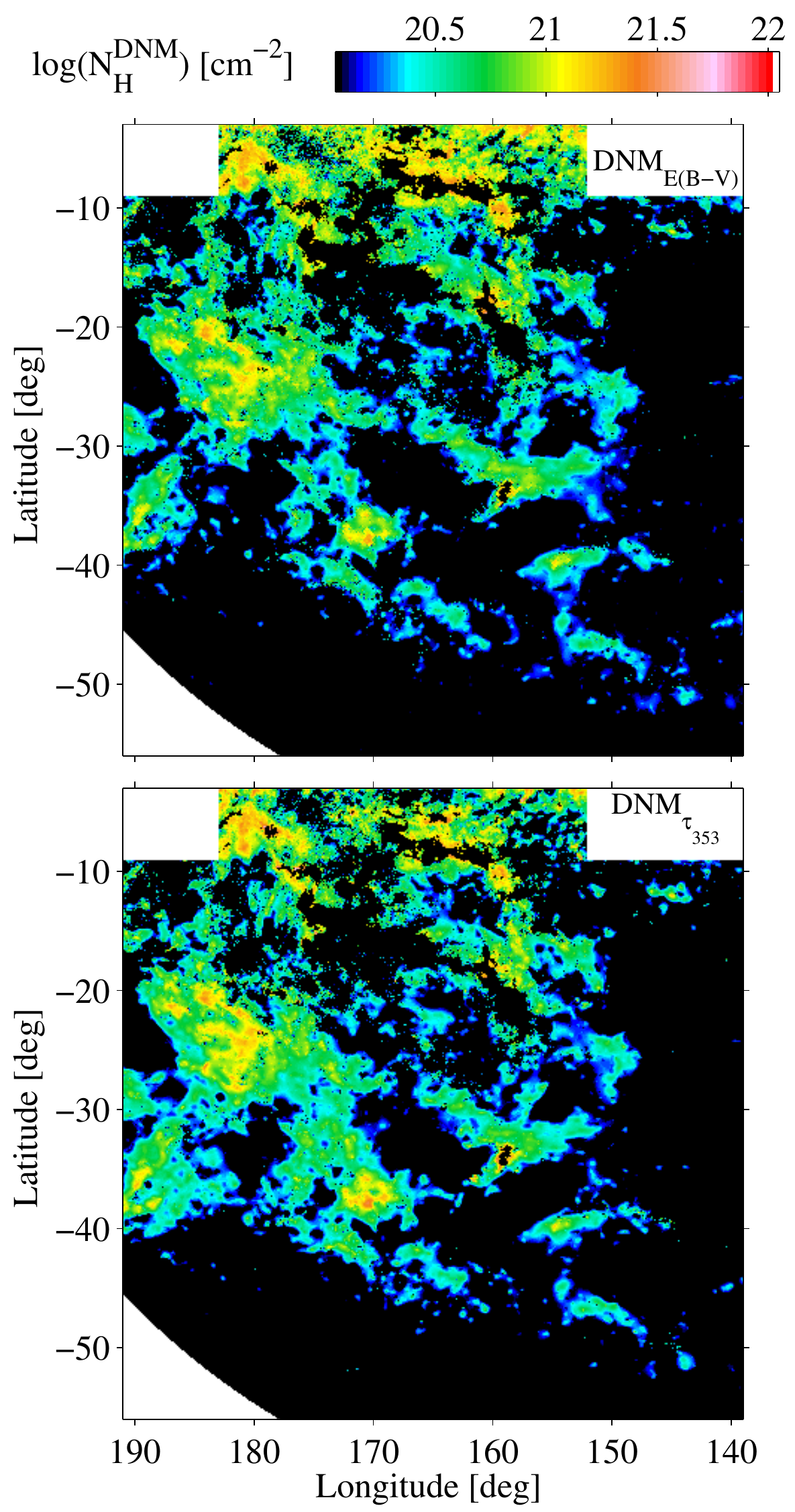}
  \caption{ Hydrogen column density maps obtained for the DNM with the \anaE (top) and \anaT (bottom) analyses of the full region (mask0).}
   \label{fig:DK_4type}
\end{figure}

\section{Results}\label{sec:res}


\subsection{Gas column density} 

The coupled analysis of \g rays and dust, with the latter in emission or reddening, takes advantage of the common underlying gas distribution in the two tracers to constrain the amount and the structure of the gas ill traced by the radio line surveys and free-free emission (\textit{i.e.}, DNM and \cosat). This gas is jointly revealed by its dust and CR content. 
We first investigate how the choice of dust tracer affects the derivation of gas column densities in the DNM and \cosat phases, then in the total gas. 

\begin{figure}
  \centering   
  \includegraphics[width=\hsize]{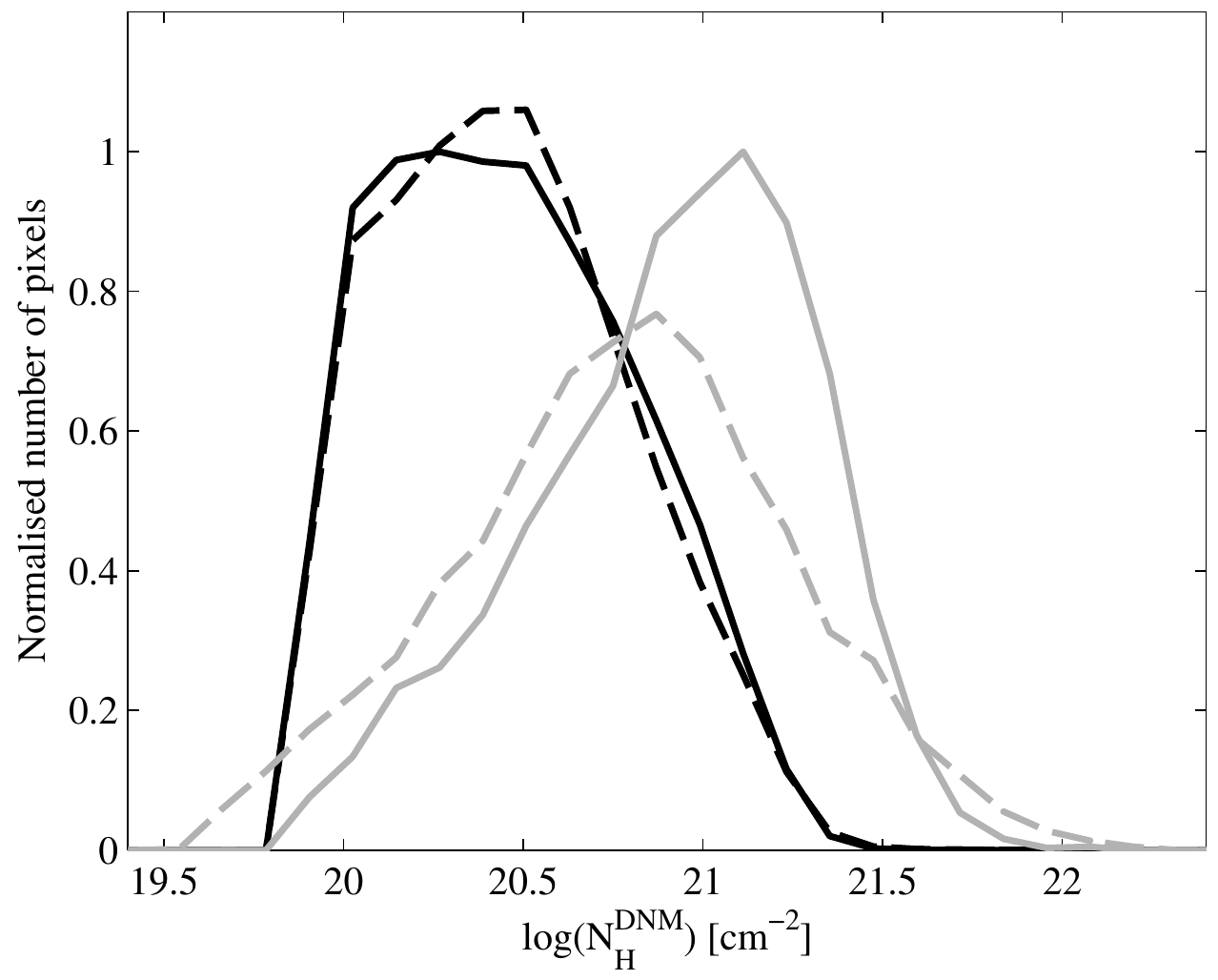}
  \caption{Number distributions of hydrogen column densities in the DNM (black) and \cosat (gray) components, obtained from the \anaT (dashed) and \anaE (solid) analyses of the full region (mask0). The numbers of pixels for each component have been normalized by the maximum value found in the \anaE analysis (solid curves)}.
   \label{fig:DK_hist}
\end{figure}

Figure \ref{fig:DK_4type} shows that the hydrogen column densities derived in the DNM from the \anaE and \anaT analyses exhibit very similar spatial structures and Fig. \ref{fig:DK_hist} illustrates that the column density values compare well across the whole range. The maps present only small differences, for instance slightly more extended diffuse cloud edges below 4$\times10^{20}$ cm$^{-2}$ in the \anaT map, and larger column densities, probably from the Galactic disc background, above 1.5$\times10^{21}$ cm$^{-2}$ near the Galactic plane ($b>-8$\degr) in the \anaE map. But the total mass in the \anaT map is only 9\% lower than in the \anaE one. 

\begin{figure}
 \vspace{0.5cm}
  \centering   
  \includegraphics[width=\hsize]{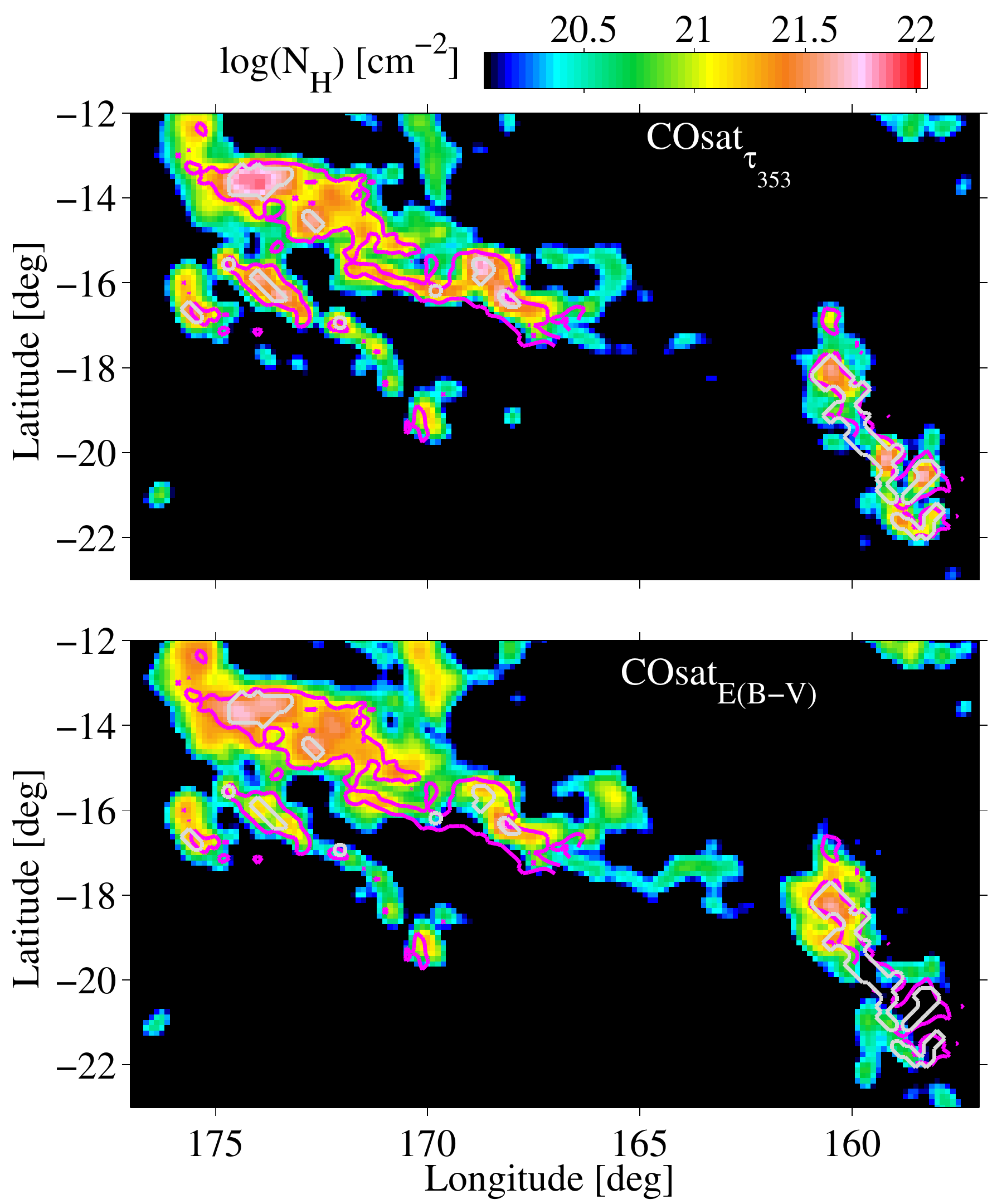}
  \caption{ Hydrogen column density map of additional gas in the $^{12}$CO-saturated parts of the Taurus and Perseus clouds, derived from the \anaT (top) and \anaE (bottom) analyses of the full region (mask0). The pixels within the grey contours have a lower angular resolution in \ebv (mask2). The magenta contours enclose regions with $^{13}$CO intensities above 2 K km s$^{-1}$ \citep{2006AJ....131.2921R,2008ApJS..177..341N}.}
   \label{fig:DNM13co}
\end{figure}

Similarly, Fig. \ref{fig:DNM13co} illustrates that the two derivations of the \cosat component present the same overall spatial structure. Yet, local differences in column densities are more accentuated in the \cosat component than in the DNM (see Figure \ref{fig:DK_hist}). The main differences arise at intermediate column densities of $(1-3)\times 10^{21}$ cm$^{-2}$ where the \anaE map yields larger values in several structures. Hence, the total \cosat mass in the \anaE map is 22\% larger than in the \anaT one. Conversely, the densest filaments and clumps emitting in the rarer $^{13}$CO isotope are enhanced in the \anaT map compared to the \anaE one (beyond  $6\times 10^{21}$ cm$^{-2}$). 
This difference results from the combination of the sharp rise  in \opa at the large column densities characterizing these cores \citep[see Fig. 16 of][]{2017A&A...601A..78R} and from the loss of contrast in reddening because of the lower angular resolution of the \ebv map due to the loss of stars in those heavily obscured directions (see the grey contours in Fig. \ref{fig:DNM13co}). 

\begin{figure*}
  \centering   
  \includegraphics[width=\hsize]{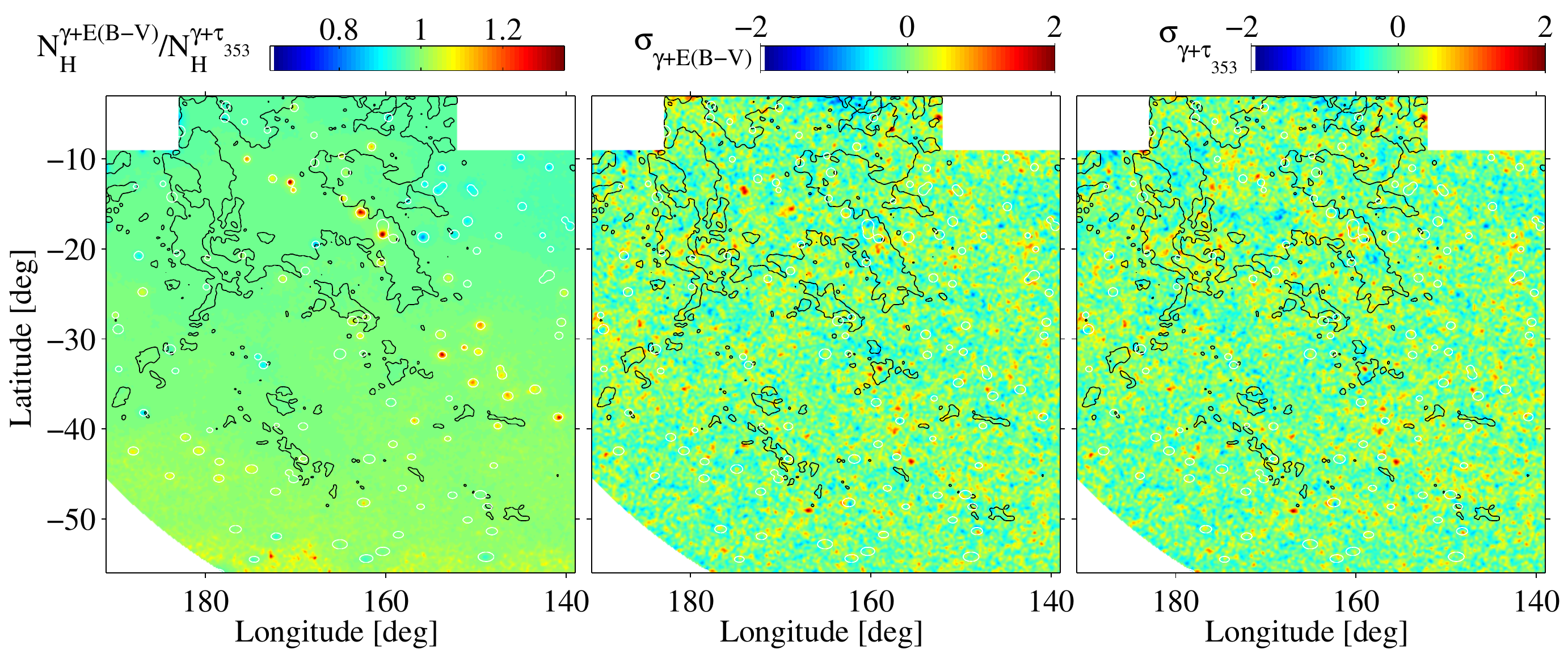}
  \caption{\textit{Left:} Map of the ratios of hydrogen column densities derived from interstellar \g rays in the \anaE and \anaT analyses, the white contours marking the positions of \g-ray point sources; \textit{Middle and right:} residual \g-ray maps (photon counts minus best-fit model, in sigma units) in the 0.4--100 GeV band for the \anaE (middle) and \anaT (right) analyses, for an \hi spin temperature of 400~K. The black contours outline the CO clouds at the 1 K km s$^{-1}$ level.}
   \label{fig:gamRES}
\end{figure*}

The \g-ray counts of gaseous origin are derived from the total data after subtraction of non-gaseous components estimated from the best-fit model (IC emission, isotropic backgrounds, and point sources). Those counts are displayed in Fig. \ref{fig:gamDust}. They are converted to total gas column density, \nhgam, using the average emissivity spectrum per gas nucleon measured in the atomic phase in the region. Figure \ref{fig:gamRES} compares the gas column densities obtained in the \anaE and \anaT analyses by showing their ratio. The subtraction of point sources results in very localized differences, by less than $\pm 30$\%, between the two gas maps. Only 10\% of the point sources present in the region have seen their flux   slightly change between the two analyses, but the differences remain compatible with the individual normalisation uncertainties given in the 3FGL catalogue (which used a different interstellar background model).

Away from those few point sources, the two column density maps agree within $\pm3\%$, except toward the most diffuse regions, at $b< -45$\degr or at $l<155$\degr and $b<-10$\degr, where differences of about $\pm10\%$ result from the noise level in the reddening map. Hence, despite the differences discussed in Fig. \ref{fig:DK_4type} and \ref{fig:DNM13co}, the \nhgam column densities are not significantly affected by the choice of dust tracer used to infer the DNM and \cosat components that complement the \hi and CO data in the \g-ray model in order to constrain the gaseous components. The robustness of the \nhgam map is due to the fact that most of the gas mass resides in the \hi-bright and CO-bright parts (which are not inferred from fits to the dust) and that the large DNM mass corresponds to intermediate column densities where the two dust tracers yield comparable structures.

Figure \ref{fig:gamRES} also compares the quality of the \g-ray fits obtained in the \anaE and \anaT analyses by showing  the residuals (photon data minus model, in sigma unit) in the overall energy band for each analysis. In both cases, the linear model provides an excellent description of the observed data in the 0.4--100 GeV band, as well as in the individual energy bands that are not shown here. The residuals remain largely consistent with noise fluctuations below $\pm 1\sigma$. The small model excess noted toward the brightest CO parts of Main Taurus, Perseus, and California clouds in the \anaT analysis \citep{2017A&A...601A..78R} disappears when using dust reddening because of the reduced contrast in the \ebv map mentioned above. This confirms that the model excess in \anaT was caused by the marked rise in \opa opacity that characterizes the cold population of grains present in the densest CO cores \citep[see Fig. 17 of][]{2017A&A...601A..78R}, rather than by a loss of CRs.

\renewcommand{\arraystretch}{1.5}
\setlength{\tabcolsep}{0.19cm}
\begin{table}
\caption{Average $E(B-V)/N_{\rm H}$ ratios, in 10$^{-22}$ mag cm$^2$, and dust opacities, \opa in 10$^{-27}$ cm$^2$, in the \hi and CO-bright gas phases of the different cloud complexes (mask0).}
\centering
\begin{tabular}{l|c|c|c|c}
 Cloud &  $\overline{\left[\frac{E({\rm B}-{\rm V})}{N_{\rm H}}\right]}_{\rm HI}$ &  $\overline{\left[\frac{E({\rm B}-{\rm V})}{N_{\rm H}}\right]}_{\rm CO}$ & \opavghi  & \opavgco  \\ \hline 
$\rm{Cet}$ & 1.61$\pm$0.01 & 2.32$\pm$0.34 & 10.3$\pm$0.1 & 17.9$\pm$4.1  \\ 
$\rm{TauS}$ & 1.96$\pm$0.01 & 2.47$\pm$0.12 & 13.7$\pm$0.1 & 18.3$\pm$1.5 \\ 
$\rm{TauN}$ & 1.70$\pm$0.01 & 2.74$\pm$0.10 & 11.0$\pm$0.1 & 18.8$\pm$1.0 \\ 
$\rm{TauM}$ & 1.89$\pm$0.03 & 2.62$\pm$0.07 & 14.2$\pm$0.1 & 27.0$\pm$1.0  \\ 
$\rm{Cal}$ & 2.11$\pm$0.01 & 2.38$\pm$0.21 & 13.7$\pm$0.2 & 17.7$\pm$1.0  \\ 
$\rm{Per}$ & 2.55$\pm$0.10 & 2.18$\pm$0.24 & 16.5$\pm$0.5 & 34.4$\pm$5.1  \\ 

\end{tabular}
\vspace{0.5cm}
\label{tab:dustNH}
\normalsize
\end{table}

\renewcommand{\arraystretch}{1.5}
\setlength{\tabcolsep}{0.16cm}
\begin{table*}
\caption{ $X_{\rm{CO}}$ factors in $10^{20}$ cm$^{-2}$K$^{-1}$km$^{-1}$s for the dust and $\gamma$-ray fits.}
\centering
\begin{tabular}{l|c|c|c|c|c|c}
 Model& Cetus & South Taurus & North  Taurus & Main Taurus & California & Perseus \\ \hline \hline
$\tau_{353}$ mask0 & 1.58$\pm$0.03 & 1.44$\pm$0.02 & 1.55$\pm$0.01 & 1.24$\pm$0.01 & 1.11$\pm$0.01 & 1.03$\pm$0.01\\ 
$\tau_{353}$ mask2 &    & 0.90$\pm$0.01 & 1.00$\pm$0.01 & 1.26$\pm$0.01 & 1.24$\pm$0.01 & 0.45$\pm$0.01\\ \hline 
$\gamma+\tau_{353}$ mask0 & 1.01$\pm$0.15 & 1.04$\pm$0.04 & 0.90$\pm$0.02 & 0.67$\pm$0.01 & 0.87$\pm$0.03 & 0.44$\pm$0.03\\ 
$\gamma+\tau_{353}$ mask2 &     & 0.79$\pm$0.05 & 0.73$\pm$0.02 & 0.70$\pm$0.01 & 0.92$\pm$0.03 & 0.27$\pm$0.03\\ \hline \hline 
$E$(B$-$V) mask0 & 1.44$\pm$0.03 & 1.06$\pm$0.01 & 1.22$\pm$0.01 & 0.97$\pm$0.01 & 0.93$\pm$0.01 & 0.49$\pm$0.01 \\ 
$E$(B$-$V) mask1 &    & 0.74$\pm$0.01 & 0.83$\pm$0.01 & 0.94$\pm$0.01 & 0.91$\pm$0.01 & 0.45$\pm$0.01 \\ 
$E$(B$-$V) mask2 &    & 0.78$\pm$0.01 & 0.87$\pm$0.01 & 0.93$\pm$0.01 & 0.96$\pm$0.01 & 0.40$\pm$0.01 \\ \hline
$\gamma$+$E$(B$-$V) mask0 & 1.00$\pm$0.14 & 0.84$\pm$0.04 & 0.76$\pm$0.02 & 0.70$\pm$0.01 & 0.82$\pm$0.03 & 0.58$\pm$0.03 \\  
$\gamma$+$E$(B$-$V) mask1 &     & 0.63$\pm$0.04 & 0.56$\pm$0.02 & 0.70$\pm$0.01 & 0.81$\pm$0.03 & 0.58$\pm$0.03 \\ 
$\gamma$+$E$(B$-$V) mask2 &     & 0.66$\pm$0.04 & 0.59$\pm$0.02 & 0.70$\pm$0.01 & 0.86$\pm$0.03 & 0.46$\pm$0.03 \\ 

\end{tabular}
\label{tab:XCOfactors}
\end{table*}

\subsection{Specific reddening and opacity variations}

Our analyses yield average dust properties per H atom in each gas phase for each cloud complex. The best-fit coefficients of the \nhi maps in the dust model directly give the average specific reddening \ebvnh (alternatively \opa opacity) in the atomic envelopes of each cloud complex. We have also used the best-fit coefficients of the \wco maps and the \g-ray estimates of the \xco factors to derive the average specific reddening (or opacity) in the CO-bright phase of each complex. The results are listed in Table \ref{tab:dustNH} together with the opacities found in the \anaT analysis \citep{2017A&A...601A..78R}. 

We have selected different masks to investigate a possible bias due to the variable resolution of \ebv map and we find only up to 10\% variation in the average \ebvnh when excluding the pixels of lower resolution. On average this effect is not significant, but it could be more important locally within the unresolved compact filaments of molecular clouds.

The average cloud opacities were found to systematically increase by 30\% to 100\% from the \hi to the CO phase. The opacity maps showed a gradual and coherent increase which was related to grain evolution from the atomic to the molecular phase \citep[see Fig. 16 and 17 of][and the references discussed in the following section]{2017A&A...601A..78R}. The present results also show a systematic increase by 30\% to 60\% in average specific reddening between the \hi and CO phases of the clouds, except for Perseus. A large fraction of this cloud is more coarsely mapped in \ebv, but the results found with mask2 also yield consistent specific reddenings in the \hi and CO phases of Perseus. However, this cloud is very compact, both in \hi and CO \citep[see Fig. 1 of][]{2017A&A...601A..78R}, and it lies behind North Taurus, so the separation between the \hi and CO contributions to the total reddening is less efficient in this cloud than in the others. 

 \begin{figure}
  \centering     
  \includegraphics[width=\hsize]{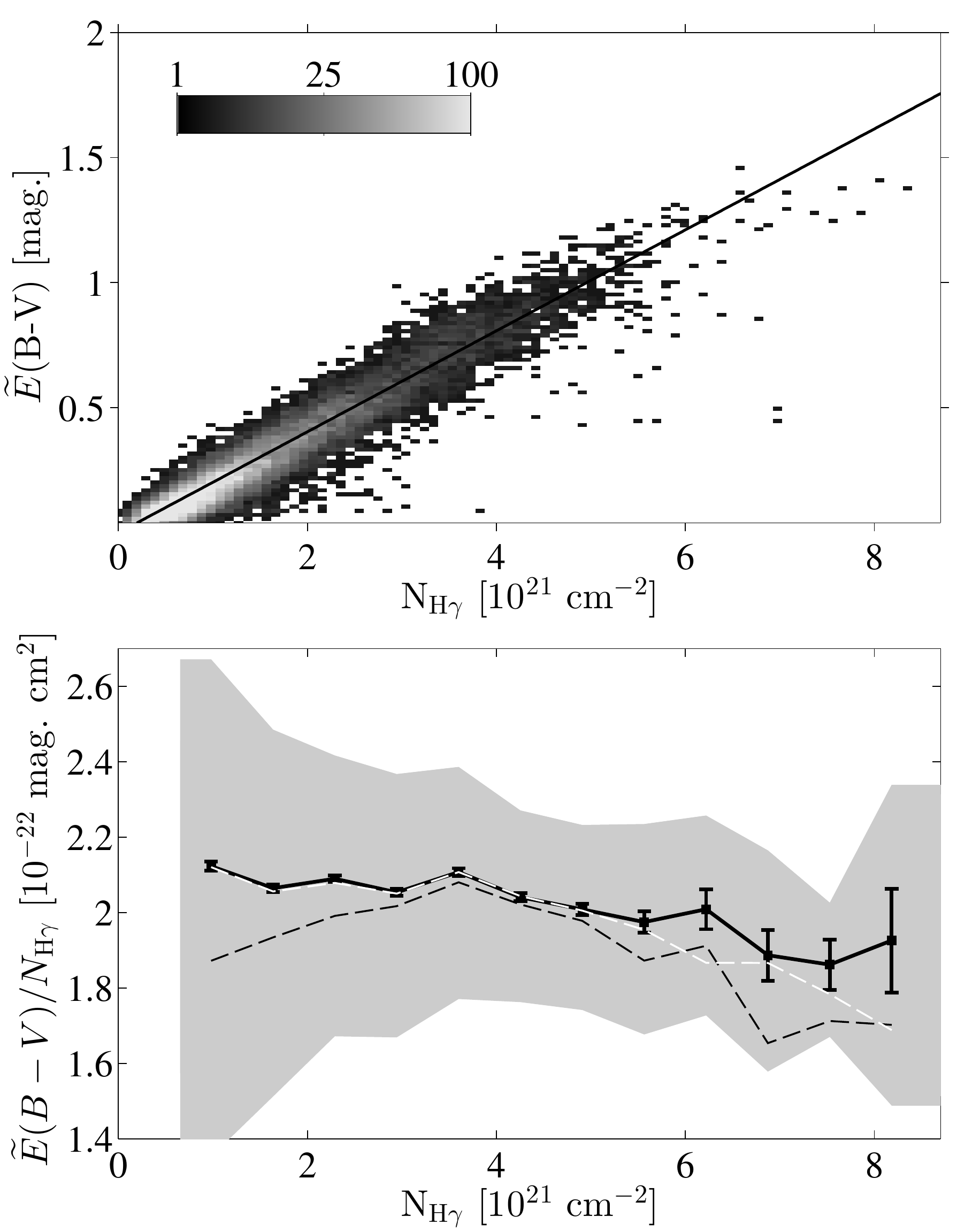}
  \caption{\textit{Upper}: 2D histogram of the correlation between the total gas column density, \nhgam, measured with the 0.4--100 GeV interstellar \g rays, and the dust reddening, $\widetilde{E}({\rm B-V})$, convolved with the LAT point-spread-function for an interstellar spectrum. The maps are sampled at a 0\fdg375 resolution. The solid line marks the best-fit slope found across the whole field (mask0), it correspond to a \ebvnhgam ratio of $(2.02\pm0.48)\times$\ebvnhunit  averaged over all gas phases. \textit{Lower}: evolution of the \ebvnhgam ratio in equally spaced bins in \nhgam. The black dashed, white dashed, and black solid curves give the means obtained with mask0, mask1, and mask2, respectively. The shaded area gives the standard deviation in each bin using mask2 and the error bars give the standard error on the means. }
   \label{fig:dustVSgam}
 \end{figure}

 \begin{figure}
  \centering 
  \includegraphics[scale=0.8]{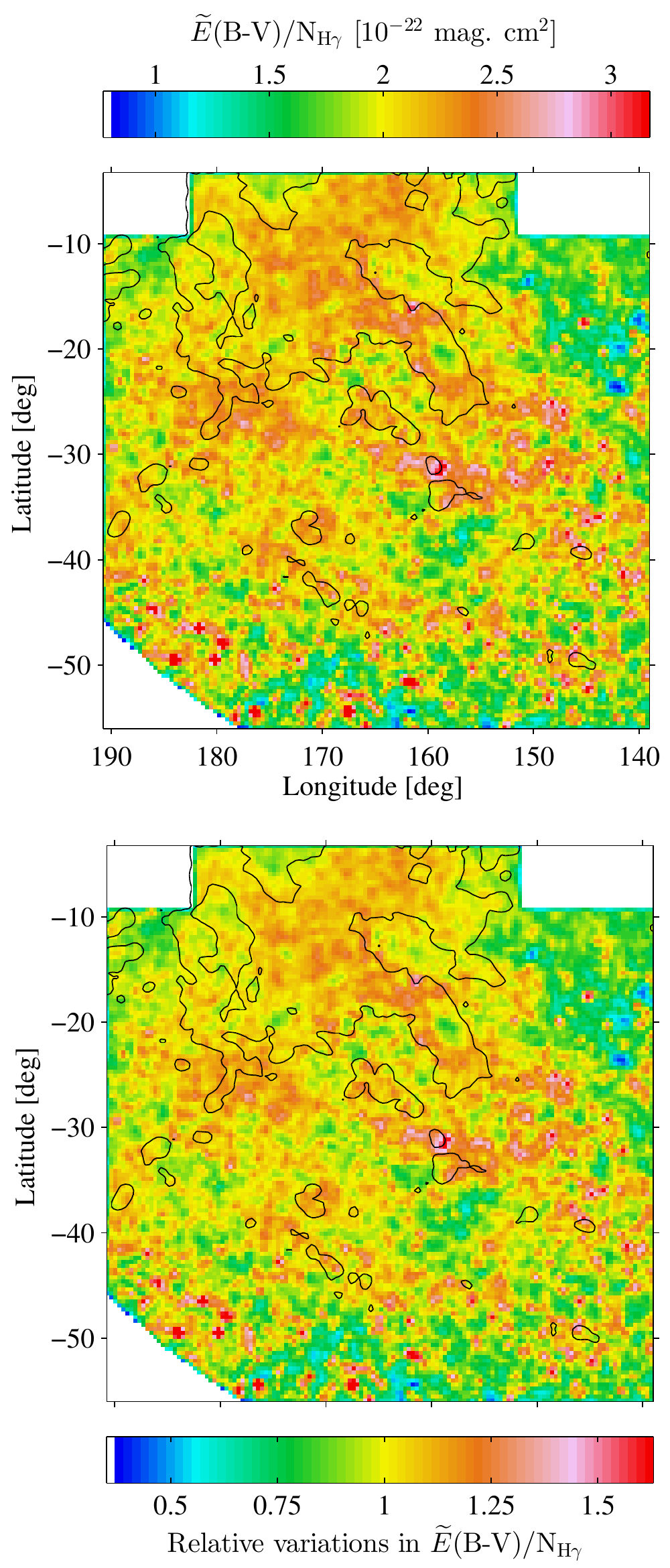}
  \caption{ Spatial variations of the amount of reddening per gas nucleon, \ebvnhgam, with the total gas measured by \nhgam at 0\fdg375 resolution (top) and its variations relative to the average value of $(2.02\pm0.48)\times$\ebvnhunit (bottom). The tilded quantities are convolved with the LAT response for an interstellar spectrum. The black contours outline the shape of the CO clouds at the level of 1 K km s$^{-1}$ in \wco intensity.}
   \label{fig:dustgamComp}
\end{figure}  

The specific reddening \ebvnh values found in the atomic parts of the North Taurus and Cetus clouds are close to the reference value of $1.7\times $\ebvnhunit of \cite{1978ApJ...224..132B} that applies to lines of sight with \ebv $< 0.5$ mag in the solar neighbourhood and in the Galactic disc. They also agree with the value of $(1.71\pm 0.07)\times$\ebvnhunit found in the atomic gas of the Taurus region from the visual extinction map of \cite{2013PASJ...65...31D} \citep{2012A&A...543A.103P}. We find, however, larger values in the range of $(1.9-2.1)\times $\ebvnhunit in the atomic gas of the Main Taurus, South Taurus, and California clouds.  \cite{2012A&A...543A.103P} had noted variations of the same magnitude between local clouds in their extinction data based on \textit{2MASS} stellar photometry. 

The average specific reddening reaches $(2.09\pm 0.15)\times $\ebvnhunit in the denser DNM at the \hi-\hd interface (given by the best-fit parameters of the \g-ray model $\overline{q_{\rm HI}}/q_{\rm DNM}$). \textit{FUSE} observations probed the same type of medium in diffuse \hd clouds up to 1 mag in reddening \citep{2009ApJS..180..125R}. The specific reddening changed between $(1.68 \pm 0.10)\times $\ebvnhunit and $(2.38 \pm 0.96)\times $\ebvnhunit depending on the inclusion or not in the fit of the data points at small reddening. The possible change in slope with reddening hinted at possible changes in dust properties with higher extinction. The average value that we find in the DNM is consistent with the \textit{FUSE} observations and the higher values we get in the range of $(2.2-2.7)\times $\ebvnhunit in the CO-bright clouds further support an evolution of the dust extinction properties.

We further note that the present \ebvnh ratios are systematically 1.2 to 2.2 times lower than the values obtained in parts of the South Taurus, Main Taurus, and Perseus clouds by \cite{2015MNRAS.448.2187C} from their visual extinction map based on \textit{2MASS} and \textit{XSTPS-GAC} stellar data. Further analyses of common regions using the same \nhgam data and direct comparisons of the \ebv and \av maps are necessary to find the origin of the discrepancy.

The good correlation between the hydrogen column density derived from \g rays and the dust reddening is presented in Fig. \ref{fig:dustVSgam} together with the evolution of the \ebv/\nhgam ratio as a function of the gas column density. The reddening map has been convolved with the appropriate LAT point-spread-function and the angular resolution of both maps has been degraded to 0\fdg375 to improve the photon statistics in the histogram bins. At this resolution, the dust reddening and gas column density linearly correlate across the whole gas range. We detect no upward inflexion in the \ebvnh ratio with increasing gas column density, contrary to the systematic rise seen in dust opacity across the same gas range \citep[see Fig. 15 of][]{2017A&A...601A..78R}.

The spatial distribution of the specific reddening is shown in Fig. \ref{fig:dustgamComp}. We also map the variations relative to the average value of $(2.02\pm0.48)\times$\ebvnhunit found for the whole field (mask0) across all gas phases. 
The \ebv/\nhgam values increase in the DNM and molecular phases compared to the diffuse regions. The 20\% drop toward Perseus is due to a relative lack of reddening, possibly due to the presence of young stellar objects rather than to an excess of gas column density. Large ratios are found in the region at $155\degr < l < 163\degr$ and $-33\degr < b < -30\degr$ where the (independent) thermal emission of the grains has a spectral index greater than 1.8 \citep[see Fig. 2 of][]{2017A&A...601A..78R}.

Across the whole region, the \ebv/\nhgam ratios vary typically by $\pm$30\% around the average value. This is far lower than the three-to-six fold variations found in the \opa ratios toward the same clouds, at the same angular resolution and with the same analysis method \citep{2017A&A...601A..78R}. The implications of this result on dust grains properties will be commented on the discussion section.


\subsection{\xco factors}\label{sec:resxco}

\begin{figure}[!h]
  \includegraphics[width=\hsize]{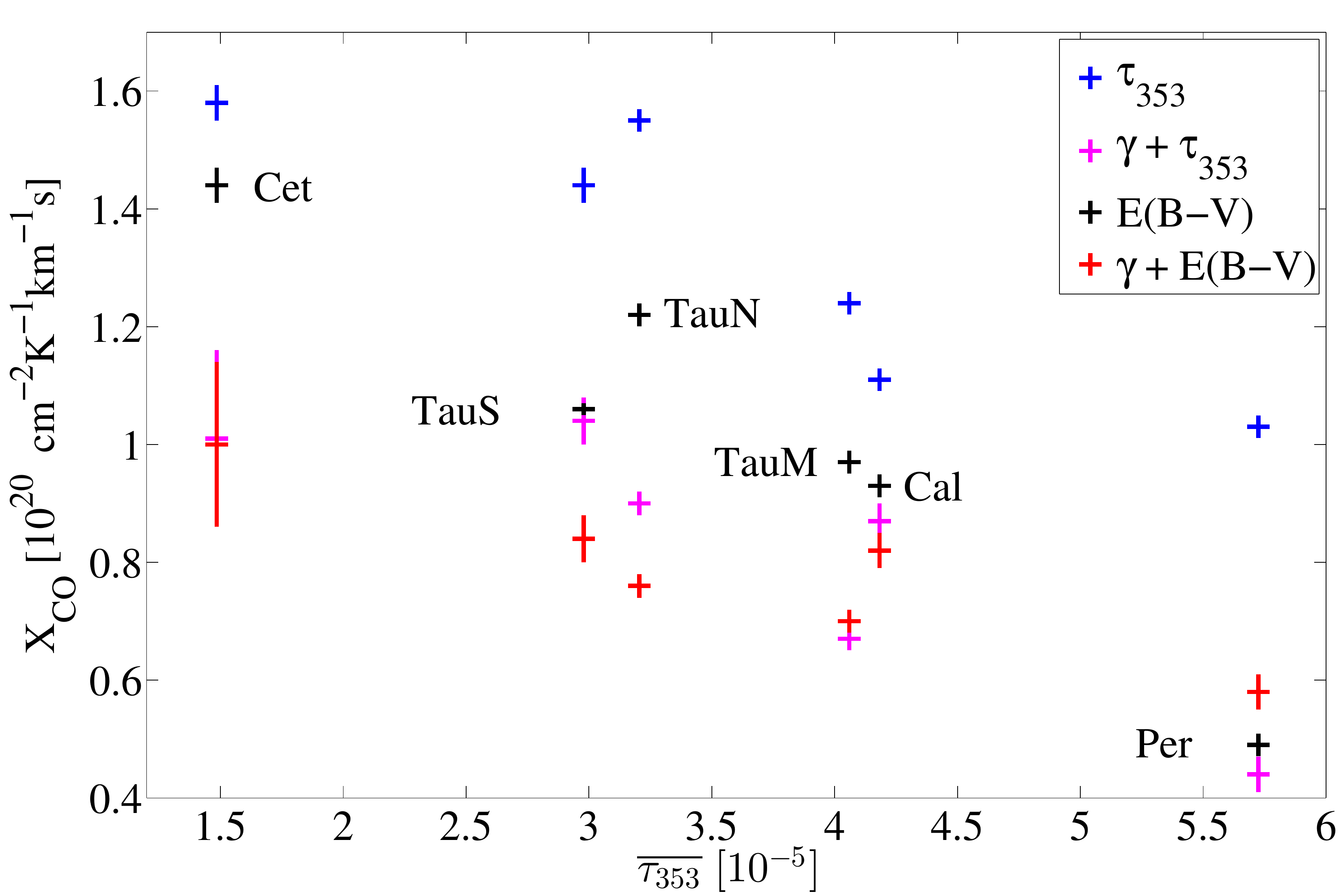}
  \caption{ Comparison of the \xco factors derived from dust and \g-ray measurements in mask0 as a function of the average dust optical depth at 353 GHz in the different clouds. The colours mark the use of different gas tracers : \ebv reddening in black, \taunu optical depth in blue, \g rays coupled to \taunu in magenta, and \g rays coupled to reddening in red.}
   \label{fig:XCOfactors_dustgam}
\end{figure}

The best-fit coefficients of the \g-ray and dust models yield independent estimates of the \xco factors under the respective assumptions of a uniform CR flux or uniform dust properties per gas nucleon in the \hi and CO phases of a given cloud complex.
The results obtained from the \anaE and \anaT analyses for different clouds and with different spatial masks are presented in Table \ref{tab:XCOfactors}. The results obtained for the whole region (mask0) are displayed in Fig. \ref{fig:XCOfactors_dustgam} as a function of the mean dust optical depth, \taunuavg, in the cloud complex.

\cite{2017A&A...601A..78R} reported significant variations in \xco when averaged over clouds that are structured differently. The \xco factors tend to decrease from diffuse to more compact clouds, as expected from the models of the formation and photodissociation of \hd and CO molecules. The models predict strong inward gradients in the relative abundance of CO and \hd, which result in a marked decline in \xco from the diffuse molecular envelopes to the dense and well-shielded cores \citep{2006MNRAS.371.1865B,2016MNRAS.455.3763B}. Hence the \xco average over a cloud should vary with the surface fraction of dense versus diffuse regions and with the degree of saturation of the CO lines in the densest parts. 
Sorting the clouds in Fig. \ref{fig:XCOfactors_dustgam} by their mean \taunuavg value partially captures their diffuse (low \taunuavg) or compact (high \taunuavg) character. The systematic decline seen in \xco with all gas tracers clearly confirms the presence of CO-to-\hd abundance gradients inside the clouds. Table \ref{tab:XCOfactors} further shows that masking out large fractions of the South Taurus and North Taurus complexes induces significant changes in their average \xco factors in all the analyses, thereby confirming the existence of spatial variations in \xco within a cloud complex because of the different states of its constituents.
The Main Taurus and California cloud complexes are only marginally affected by the change in mask. This is why their average \xco factors remain stable for the different masks. 

The \g-ray values of the \xco factor in the Perseus cloud should be taken with care because of the significant level of cross-correlation between the compact \hi and CO phases of the cloud when convolved at the angular resolution of the LAT, especially at low energies \citep[see Sec. 6.1 of][]{2017A&A...601A..78R}. This is even more true when further lowering the dynamic ranges by removing pixels in mask2. The \xco factors in Perseus vary significantly when masking out the densest gas pixels: the \taunu-inferred value decreases by a factor of 2.3 because the masked pixels correspond to regions where the grains have particularly large emission opacities; the change is still pronounced in \g rays, but reduced to a factor of 1.6, in the \anaT analysis because the masked pixels have large column densities in the \taunu-inferred \cosat component; the change is reduced to 20\% in the \g-ray fit coupled to the dust reddening. Based on the \g-ray and reddening results, the average \xco factor in Perseus should be close to (0.5-0.6) $\times$ \xcounit. 


By construction of the models, the DNM gas component has negligible influence on the \xco estimates \citep{2015A&A...582A..31A}. Figure \ref{fig:XCOfactors_dustgam} further shows that coupling the \g-ray analysis to different dust tracers to infer the \cosat components does not change much the \g-ray values of \xco. They remain stable in the Cetus, Main Taurus, and California clouds and decrease by 20\% in the North and South Taurus clouds when replacing dust emission by reddening. The figure also illustrates that the reddening-based measurements are systematically closer to the \g-ray ones than the measurements based on dust emission, but they still differ by 15\% to 60\%. \cite{2017A&A...601A..78R} noted that the \g-ray and \taunu estimates of \xco differed more at larger \taunuavg because compact clouds exhibit large gas column densities where the dust grains have particularly large opacities, but we do not find a similar divergence between the \g-ray and reddening estimates of \xco. The reddening-based factors are systematically smaller than those obtained with dust in emission (see Fig. \ref{fig:XCOfactors_dustgam}), so the results corroborate the existence of a strong opacity bias in \xco measurements when they are based on dust emission only. 



\section{Discussion and conclusions}\label{sec:ccl}

Several studies have reported variations in dust opacity \cite[e.g.,][]{2003A&A...398..551S,2009ApJ...701.1450F,2012ApJ...751...28M,2013A&A...559A.133Y,2013ApJ...763...55R,2014A&A...571A..11P,2014A&A...566A..55P,2015A&A...579A..15K,2017A&A...601A..78R}. The extinction curves also respond sensitively to local conditions and contain unique information about the grains along the lines of sight \citep{2007ApJ...663..320F}.  Changes in the extinction law have also been reported \citep{1989ApJ...345..245C,2013MNRAS.428.1606F,2016ApJ...821...78S}. In both cases, the changes have been attributed to the evolution of grain properties under the local conditions. Irradiation can modify their chemical composition and structure while mantle accretion, coagulation, and ice coating change their structure.
In Sec. \ref{sec:def}, we have recalled how the \opa opacity and the \ebvnh specific reddening respond to several grain properties in emission and in extinction. The results presented here and by \cite{2017A&A...601A..78R} show that both the emission and extinction characteristics significantly vary across the anti-centre clouds, but not in strict spatial correlation and not with the same amplitude. In the following we discuss how the variations relate with the spectral characteristics of the dust emission and how they compare with predictions from recent theoretical models \citep{2013A&A...558A..62J,2015A&A...579A..15K}. 

Figure \ref{fig:ebvnh_beta_T} displays how the dust characteristics vary with the colour temperature, $T_{\rm dust}$, and spectral index, $\beta$, of the thermal emission of the large grains in the 353–3000 GHz band \citep{2014A&A...571A..11P}. The clouds have been sampled at an angular resolution of 0\fdg375 which corresponds to a linear scale of 1 to 3 pc in the different complexes. The upper plots map the changes in specific reddening and in opacity in absolute values, while the third plot shows the relative changes in the \ebvtau ratio to illustrate how the extinction and emission properties evolve relative to each other. To show the amplitude of the relative variations, we have divided the \ebvtau ratio by the mean slope of $1.49 \times 10^4$ mag found in the correlation between \taunu optical depths and quasar colour excesses across the sky, mostly through atomic clouds \citep{2014A&A...571A..11P}. 
The bottom plot shows the mean gas column density inferred in \g rays across the range of dust spectra in order to link the dust properties with their gaseous environment, from the diffuse clouds where the grains are immersed in low-density and mostly atomic gas and where they are well exposed to the ISRF (at low $\beta$ index and large $T_{\rm dust}$ temperature), up to the dense, molecular, and obscured media in our sample. 

Figure \ref{fig:ebvnh_beta_T} highlights that the dust properties per gas nucleon in emission and in extinction vary coherently across the ISM phases, but that the gradients in specific reddening and in opacity are not commensurate. The rise in opacity, is more than twice as large as the coincident rise in specific reddening, as the gas becomes denser. We discuss these evolutions in the diffuse and dense media in turn.

\subsection{Dust evolution in the diffuse medium}

In the \anaT analysis of the nearby anti-centre clouds \citep{2017A&A...601A..78R}, we have showed that the emission opacity \opa increases by a factor of two in the \hi and DNM phases above the diffuse-ISM value of $(7.0 \pm 2.0)\times$\opaunit found in tenuous high-latitude cirruses \citep{2014A&A...566A..55P,2014A&A...571A..11P}. 
Comparable variations have been noted in the atomic gas across the sky, with values ranging from 6.6 to about $11\times$\opaunit \citep{2014A&A...571A..11P}. Using the Planck optical depth at 353 GHz and \nhi column densities from GALFA, \cite{2015ApJ...811..118R} also reported up to three-fold variations in the diffuse ISM.

\begin{figure}
  \centering 
  \includegraphics[width=\hsize]{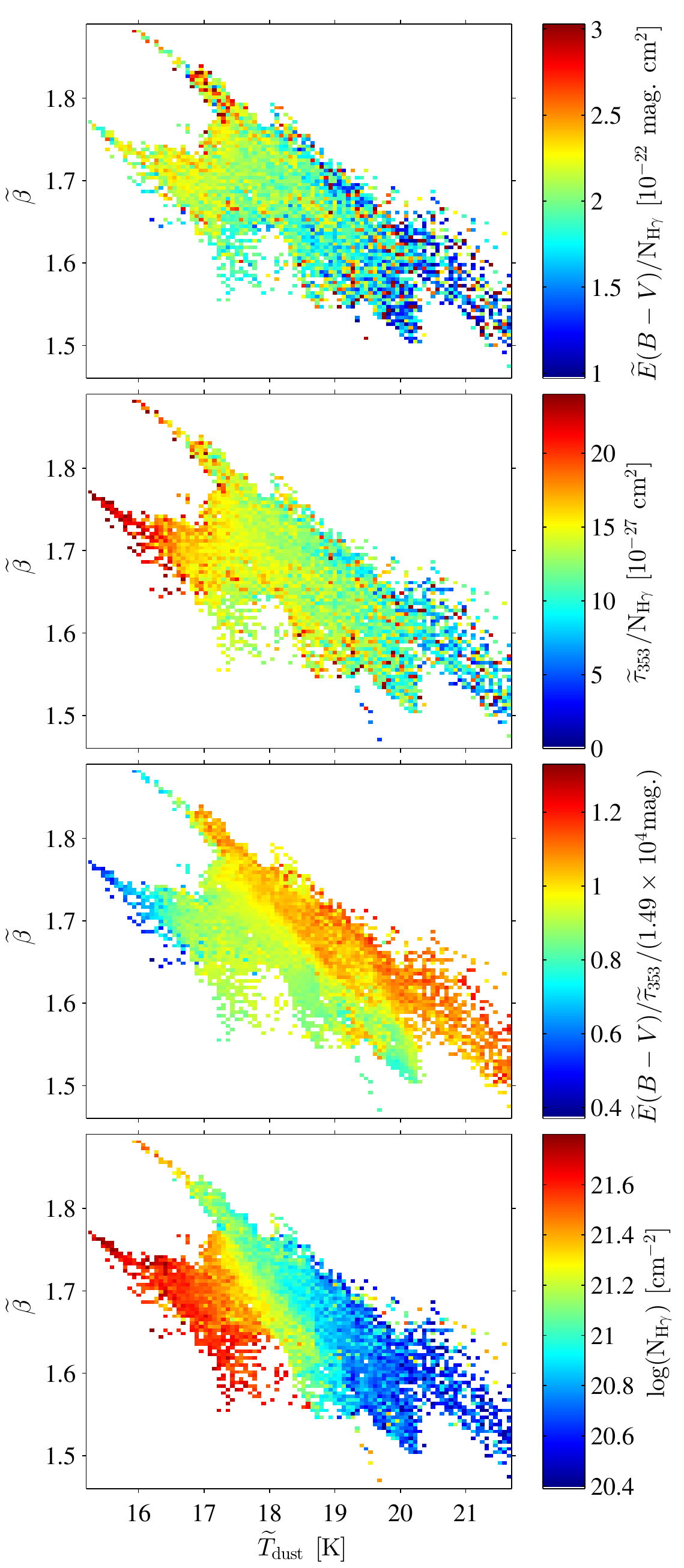}
  \caption{\small Evolution of the \ebvnhgam specific reddening (top), $\widetilde{\tau}_{353}/$\nhgam opacity (second), relative variation in the $\widetilde{E}({\rm B-V})/\widetilde{\tau}_{353}$ ratio (third), and in total gas column density  \nhgam derived from interstellar \g rays (bottom), in 0\fdg375-wide bins, as a function of the colour temperature, $T_{\rm dust}$, and spectral index, $\beta$, of the thermal dust emission. The tilde quantities have been convolved with the LAT response for an interstellar spectrum. The variations in the $\widetilde{E}({\rm B-V})/\widetilde{\tau}_{353}$ ratio are given with respect to the mean value of $1.49 \times 10^4$ mag found from quasar colours through diffuse atomic clouds \citep{2014A&A...571A..11P}.}
   \label{fig:ebvnh_beta_T}
 \end{figure}

In the anti-centre region, we find that the specific reddening increases by less than 50\% from the very diffuse gas to the denser \hi and DNM parts of the clouds. The values remain close to the reference of 1.7$\times 10^{22}$ cm$^2$ mag of \cite{1978ApJ...224..132B} toward the sight lines intercepting atomic gas. They increase to an average of  $(2.09\pm0.15)\times$\ebvnhunit in the DNM, but we find no clear trend with \nh across the \hi and DNM phases (see Table \ref{tab:dustNH} and Fig. \ref{fig:dustgamComp}), probably because they span the same range of column densities even though the UV field and the gas chemistry differ between the two media. The \ebvnh values seen below 1.5$\times 10^{22}$ cm$^2$ mag in Fig. \ref{fig:ebvnh_beta_T} and \ref{fig:dustgamComp} correspond to hardly reddened regions ($< 0.08$ mag) where the difficult reddening measurements should taken with care. Hence, the amplitude of the opacity rise in the \hi and DNM phases significantly contrasts with the small variations in \ebvnh ratio. 


Figure \ref{fig:ebvnh_beta_T} shows that the increase in emission opacity and in specific reddening both relate to an increase in $\beta$ index and a decrease in colour temperature. Recent theoretical studies \citep[e.g.][]{2015A&A...577A.110Y,2015A&A...579A..15K,2016A&A...588A..44Y} suggest that the spectral changes in $\beta$ relate to gas density-dependent processes and that the variations in opacity, $\beta$ index, and colour temperature are closely connected. According to \cite{2017A&A...602A..46J}, 
the carbon and silicate grains in the diffuse ISM have H-poor, aromatic-rich carbonaceous mantles. Their optical properties correspond to the low $\beta$, large temperature ($\gtrsim 20$~K), and low opacity part in Fig. \ref{fig:ebvnh_beta_T}b. By accretion of C and H atoms from the gas phase, an additional H-rich amorphous carbon mantle forms on the surface of grains. It is aromatised in regions well exposed to UV radiation, and more aliphatic in attenuated regions. The change in thickness and composition of the mantle induces a large increase by typically 0.3 in $\beta$ index, as well as a doubling of the emission opacity (emissivity), so the grain equilibrium temperature drops by ${\sim}2$~K.
The changes in opacities and temperatures found in the anti-centre clouds suggest that the accretion and evolution of the additional mantle would occur primarily in the dense \hi and DNM phases, which correspond in Fig. \ref{fig:ebvnh_beta_T}d to the green-to-orange zones and to the red zone at $\beta < 1.65$ and $17 < T_{\rm dust} < 18.5$~K). The amplitudes of the variations in \opa and $T_{\rm dust}$ are compatible with the model predictions, but for a smaller change in $\beta$ index in the observations.

At UV and optical wavelengths, the extinction efficiency factor (defined as the sum of the absorption and scattering efficiency) tends to the limit $Q^{\rm ext}=Q^{\rm abs}+Q^{\rm sca}=2$ for grain sizes larger than \mbox{0.1 $\mu$m} \citep{1983asls.book.....B,1990A&A...237..215D}. This condition is verified in the model of \cite{2015A&A...579A..15K} as they find $Q^{\rm abs}\sim Q^{\rm sca}\sim 1$ up to 0.6 $\mu$m in wavelength for all the grains larger than 0.11 $\mu$m and the aggregates larger than \mbox{0.17 $\mu$m}. Mantle accretion does not significantly change the size of the grains. It induces only a 10$\%$ increase in $R_{\rm V}$, from $3.5 \pm 0.2$ to $3.9 \pm 0.2$, and a less than 20\% increase in extinction efficiency at B and V wavelengths \citep{2015A&A...579A..15K}. Such constraints are compatible with the small dispersion in \ebvnh found in the atomic and DNM phases of the anti-centre clouds. Deviations by $\pm 30$\% suggest that the population of large grains is rather stable in size distribution, absorption cross section, and in proportion to the gas mass across those phases (see Eq. \ref{eq:ebvnh}).
Thus the three-to-six fold increase in emission opacity (Eq. \ref{eq:taunh}) would primarily be due to a change in grain emissivity $\kappa_\nu$ caused by mantle accretion in the dense \hi and diffuse \hd media, rather than to an increase in dust-to-gas mass ratio, $R_{\rm DG}$.  

\cite{2016ApJ...821...78S} have selected 37,000 stars with stellar parameters estimated from the APOGEE spectroscopic survey and they have used \mbox{\textit{Pan-STARRS} 1} photometry to constrain the extinction curve from optical to infrared wavelengths and to derive the $R_{\rm V}$ factor. Up to $\sim2$ mag in reddening their values of $R_{\rm V}$ vary in the 3.1-3.5 range around a mean of about 3.3 which is lower than the model prediction. The observed factors do not systematically increase with reddening in this range. The amplitude of these $R_{\rm V}$ variations is consistent with the dispersion we measure in \ebvnh in the diffuse anti-centre clouds.
\cite{2016ApJ...821...78S} further note that larger $R_{\rm V}$ values relate to lower $\beta$ indices, at variance with the model of \cite{2015A&A...579A..15K}, but compatible with the decrease in \ebvnh and \ebvtau we see in Fig. \ref{fig:ebvnh_beta_T} toward the lowest $\beta$ indices and largest temperatures.
In the proximity of UV sources or at low densities where there is less screening against UV radiation, the grains are more exposed to the ISRF which causes their darkening and an increase in $R_{\rm V}$ with time \citep{2010MNRAS.408..535C}. The \ebvtau ratio is also expected to decrease as the ambient radiation field hardens \citep{2015A&A...577A.110Y}. Figure 16 of \cite{2017A&A...601A..78R} shows how the dust specific power (radiance per gas nucleon) varies across the anti-centre region. For grains in thermal equilibrium, the specific power traces variations in the heating rate. Away from localized hot spots associated with \hii regions, the relation between the specific power and specific reddening is unclear. Their variations in the \hi phase correlate at small spatial scales in places and not in others. 

\subsection{Dust evolution in the molecular regions}

Dust evolution continues in the denser ISM phases. The very small carbonaceous grains rapidly coagulate onto the large ones at gas volume densities above a few $10^3$~cm$^{-3}$ that are typical of CO-bright media \citep{2017A&A...602A..46J}. \cite{2015A&A...579A..15K} predict only a small impact on the $\beta$ index in the 353--3000 GHz band, which should increase by less than 0.1 whether or not the final grains get coated with ice. The structural change and ${\sim}50$\% increase in emission opacity before and after coagulation should lower the grain temperatures below 18~K if the aggregates are ice free. In the case of an ice mantle, the grains radiate much more efficiently (by a factor of 3.5) and they cool below 17~K. The overall change in opacity between the diffuse atomic and dense molecular media is therefore predicted to be of order 3 without ice coating and 7 with ice \citep{2015A&A...579A..15K}. 

Opacity values that are 2 to 3 times above the diffuse-ISM average have been found in the local ISM \citep{2009ApJ...701.1450F,2011A&A...536A..25P,2013A&A...559A.133Y,2003A&A...398..551S}. Steeper rises have been found in nearby clouds, with factors of 3 to 6 in the Taurus and Perseus clouds  \citep{2017A&A...601A..78R}, and 2 to 4.6 in the Chamaeleon clouds \citep{2015A&A...582A..31A}. Variations of this magnitude agree with the above predictions. Yet, the opacity increase in the more massive California molecular cloud is less than 2-fold \citep{2017A&A...601A..78R}. Whether this disparity could be due to the loss of small-scale contrast at larger distance or to a different dust state remains to be verified.
Large opacities are associated in Fig. \ref{fig:ebvnh_beta_T} with spectral indices $\beta> 1.65$ and colour temperatures below 17~K in reasonable agreement with the model predictions for coagulation. The largest opacities in the compact filaments of Taurus (highest column densities in Fig. \ref{fig:ebvnh_beta_T}d) would then be due to ice-coated aggregates with a 0.1 lower $\beta$ index than in the model of \cite{2015A&A...579A..15K}.  Directions with large extinctions (\av>3.3 mag) in the Taurus clouds are indeed associated with the detection of water ice \citep{1988MNRAS.233..321W}.

\cite{2015A&A...579A..15K} use aggregates that are about three times larger than the size of the constituting grains (at the peak of the mass-size distribution, see Fig 1. of their paper). Similarly, \cite{2013A&A...559A.133Y} have observed a two-fold increase in opacity at 250 $\mu$m toward a molecular filament in the Taurus cloud and they have modelled it as the result of grain growth, by an average factor of five. When aggregates are formed, \cite{2015A&A...579A..15K} predict a larger $R_{\rm V}$ value of $4.9 \pm 0.2$ independent of ice coating. The $R_{\rm V}$ factors measured by \cite{2016ApJ...821...78S} start to increase with increasing reddening for \ebv $>2$ mag, but they hardly reach values beyond 4 (only 1\% of the sample), possibly because the survey does not point specifically to dense molecular regions, but scans larger scales than the compact CO filaments.
\cite{2013MNRAS.428.1606F} have used deep red-optical data from Megacam on the MMT\footnote{\url{https://www.mmto.org/}}, combined with near-infrared data from UKIDSS, to probe the extinction law in two 1\degr--size sub-structures of the Perseus cloud, known as B5 in the eastern end (at $l=$160\fdg6 and $b=-$16\fdg8) and L14484-L1451-L1455 in the western end (at $ l=$158\fdg3, $b=-$21\fdg5).
They find that the extinction law changes from the diffuse value of $R_{\rm V}\sim3$ at \av$=2$ mag to a dense-cloud value of $R_{\rm V}\sim5$ at \av$=10$ mag in both sub-structures. These estimates compare with the theoretical value for aggregates and suggest changes in $R_{\rm V}$ from 3 to 5 with increasing gas density in the CO-bright phase.

The combination of grain growth in size and of lower extinction in the optical band (increased $R_{\rm V}$) should lead to a marked decline in specific reddening from the \hi to CO media (see eq. \ref{eq:ebvnh}). In the anti-centre clouds, we find a gradual increase rather than a decrease of the mean \ebvnh ratio, from $(1.84\pm0.10)\times$\ebvnhunit in the \hi phase, to $(2.09\pm0.15)\times$\ebvnhunit in the DNM, and $(2.53\pm0.12)\times$\ebvnhunit in the CO-bright phase. Furthermore, \mbox{Fig. \ref{fig:dustgamComp}} shows that the changes in \ebvnh do not anti-correlate with the \nh map toward CO clouds, except toward Perseus.
Inside the latter, the specific reddening is 30$\%$ lower than the average in the western end and 10$\%$ to 20$\%$ larger in the eastern end, whereas \cite{2013MNRAS.428.1606F} find the same increase in $R_{\rm V}$ in both regions. In the other anti-centre CO clouds, the relative stability of the observed \ebvnh ratio may hint at a significant increase in gas-to-dust mass ratio and/or in $Q^{\rm ext}$ to compensate for the factor of 4.8 increase due to grain growth and $R_{\rm V}$ increase.
In their study of the Magellanic Clouds, \cite{2014ApJ...797...86R} found that depletion measurements suggest that gas-phase metals may accrete onto dust grains in molecular clouds, thus changing the dust-to-gas mass ratio, $R_{\rm DG}$. Additionally $R_{\rm DG}$ variations may result from the turbulent clustering of dust grains in molecular clouds. \cite{2016MNRAS.456.4174H} have simulated the behaviour of aerodynamic grains in highly supersonic, magnetohydrodynamic turbulence typical of molecular clouds. According to their simulation, the dust can form highly filamentary structures with a characteristic width much thinner than that of gas filaments, or sometimes the dust filaments form in different locations than the gas filaments. These differences result in localized variations of large amplitude in the dust-to-gas mass ratio. Yet, such variations cannot explain the gradual decline in \ebvtau as the gas becomes denser (see eq. \ref{eq:ebvtau} and Fig. \ref{fig:ebvnh_beta_T}c), so intrinsic changes in the absorption and scattering properties of the grain aggregates in the optical bands must play a role in maintaining high specific reddening values in the dense CO clouds.

Alternatively, the lack of a significant decline in specific reddening with increasing \hd column density may be due to limitations in the determination of the reddening in obscured regions where few stars are available to sample the extinction.
Extinction estimates are derived from magnitude-limited stellar catalogues and therefore suffer from a bias to lower values because the most heavily extinguished stars are not detected  \citep{2015MNRAS.452.2960S}. This is particularly true for lines of sight with higher column densities and it could partially explain the lack of a significant decline in the \ebvnh ratio.
Moreover, the compact molecular filaments associated with ice mantle formation are not fully resolved in \g rays, nor in reddening (see Fig \ref{fig:DNM13co} where $^{13}$CO emission coincides with the lower resolution pixels). So the actual variations in \ebvnh may be steeper than presently observed.

Figure \ref{fig:ebvnh_beta_T} highlights the complex variations of dust properties and dust spectra with environment. Figures \ref{fig:XCOfactors_dustgam} and \ref{fig:ebvnh_beta_T} as well as previous studies \citep{2014A&A...571A..11P,2015A&A...582A..31A} indicate that grain evolution severely limits the linear gas-tracing capability of dust emission to the \hi phase unless we can understand and model the grain emissivity changes. The extinction properties of the large grains have been shown here to be less variable, though still not uniform across the ISM phases.
In this context, it is essential to take advantage of the total gas tracing capability of the \g rays produced by cosmic rays to quantify \opa and \ebvnh gradients in other nearby clouds, with both larger and smaller mass than those present in the anti-centre sample, in order to confirm the trends and relative amplitudes of the environmental changes in dust properties. The comprehensive photometry data on stars and quasars from future deep synoptic surveys, such as LSST \footnote{\url{https://www.lsst.org/scientists/scibook}}, will also clarify the robustness of the reddening measurements toward the densest molecular cores.
\begin{acknowledgements}
The \textit{Fermi} LAT Collaboration acknowledges generous ongoing support
from a number of agencies and institutes that have supported both the
development and the operation of the LAT as well as scientific data analysis.
These include the National Aeronautics and Space Administration and the
Department of Energy in the United States, the Commissariat \`a l'Energie Atomique
and the Centre National de la Recherche Scientifique / Institut National de Physique
Nucl\'eaire et de Physique des Particules in France, the Agenzia Spaziale Italiana
and the Istituto Nazionale di Fisica Nucleare in Italy, the Ministry of Education,
Culture, Sports, Science and Technology (MEXT), High Energy Accelerator Research
Organization (KEK) and Japan Aerospace Exploration Agency (JAXA) in Japan, and
the K.~A.~Wallenberg Foundation, the Swedish Research Council and the
Swedish National Space Board in Sweden. Additional support for science analysis during the operations phase is gratefully acknowledged from the Istituto Nazionale di Astrofisica in Italy and the Centre National d'\'Etudes Spatiales in France. This work performed in part under DOE
Contract DE-AC02-76SF00515.
The authors acknowledge the support of the Agence Nationale de la Recherche (ANR) under award number STILISM ANR-12-BS05-0016. 
\end{acknowledgements}

\bibliographystyle{aa}
\bibliography{biblioG3}

\appendix

\section{Choice of the dust extinction map for the analysis} \label{sec:AVcomp}
This section presents a comparison of several extinction maps and a study of their correlation with the dust optical depth at 353 GHz from \cite{2014A&A...571A..11P} and with the hydrogen column density derived from interstellar \g rays in our previous analysis \citep{2017A&A...601A..78R}. The indications given by this comparison have guided the choice of the extinction map to be used with the \g rays to study the \ebvnh ratios in the different gas phases.

The main purpose of this preliminary comparison was to quantify the differences between dust extinctions derived with several methods and survey data, with a particular consideration of how the distance-extinction degeneracy affects 2D maps compared to 3D versions, and on how the stellar statistics in the optical and near infrared surveys affect the results.

\begin{figure*}
  \centering               
  \includegraphics[width=\hsize]{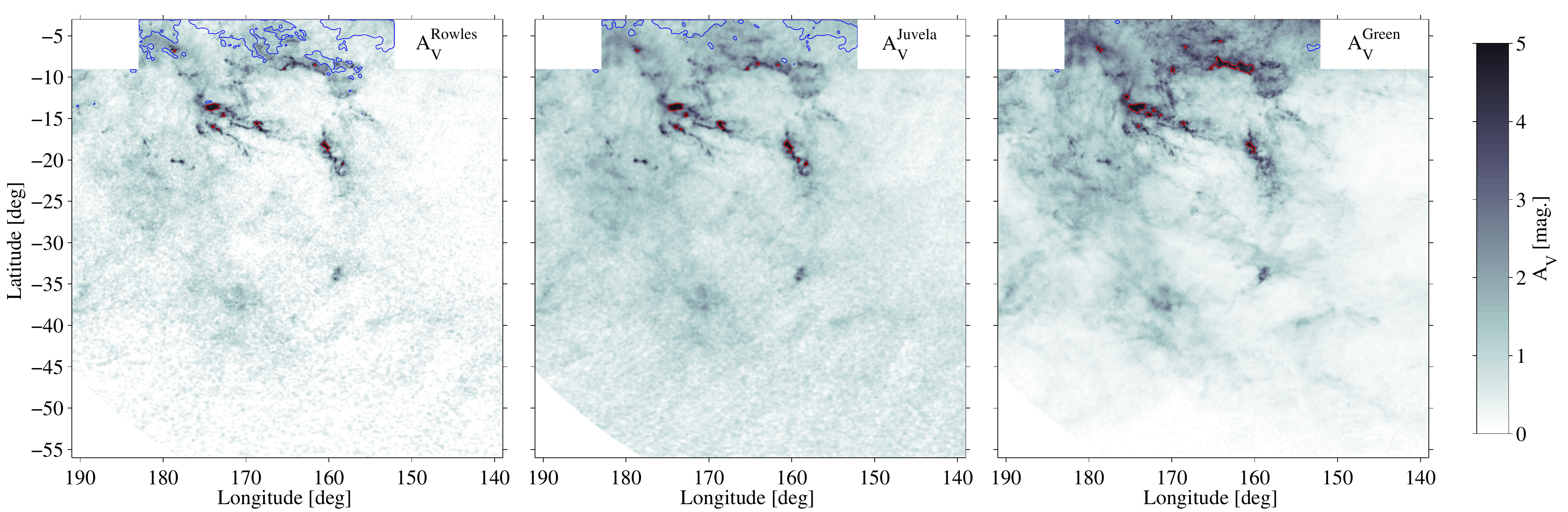}
  \caption{Visual extinction maps derived from \cite{2009MNRAS.395.1640R,2016A&A...585A..38J,2015ApJ...810...25G} (from left to right) with the extinction law of \cite{1989ApJ...345..245C}. The blue and red contours correspond to pixel selections at $0<$ \av$<1.6$ mag and at \av$>4$ mag, respectively, in regions where \nh $> 4.5 \times 10^{21}$ cm$^{-2}$. These selections in \av and \nh are shown as the blue and red areas in Fig. \ref{fig:AVnhtau}. The (\textit{l,b}) directions that do not match both conditions have been masked to generate the contours that encompass the selected pixels. 
}  
  \label{fig:AVmaps}
\end{figure*}
\begin{figure*}
  \centering               
  \includegraphics[width=\hsize]{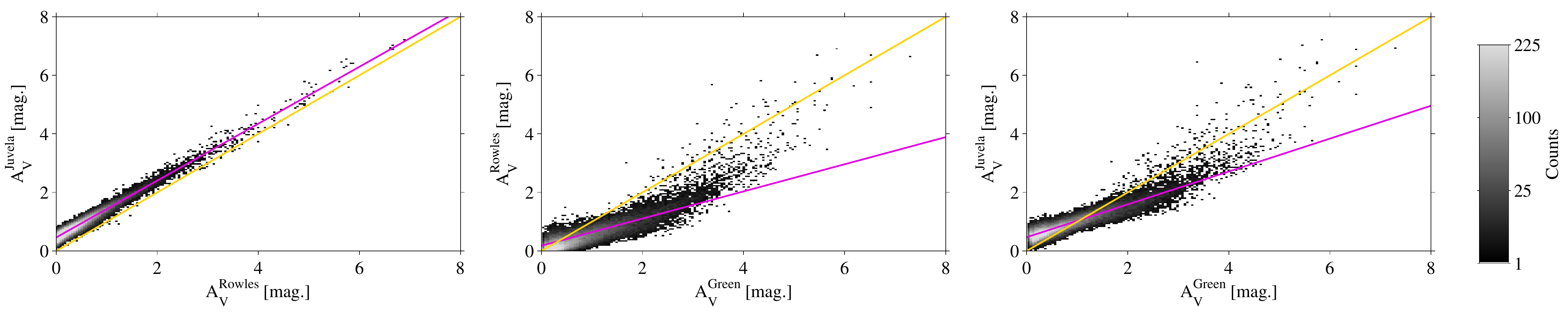}
  \caption{\textit{Top}: 2D histograms of the correlations between the visual extinction maps derived from \cite{2009MNRAS.395.1640R}, \cite{2016A&A...585A..38J}, and \cite{2015ApJ...810...25G} with the extinction law of \cite{1989ApJ...345..245C} ($R_{\rm V}$=3.1, $A_{\rm{V}}/A_{\rm{J}}$=3.55).
All the maps are compared at a common angular resolution in 0\fdg25${\times}$0\fdg25 bins. The yellow lines correspond to a one-to-one correlation. In each panel the magenta line corresponds to the best-fit of a linear regression.} 
  \label{fig:AVcomp}
\end{figure*}
\begin{figure*}
  \centering               
  \includegraphics[width=\hsize]{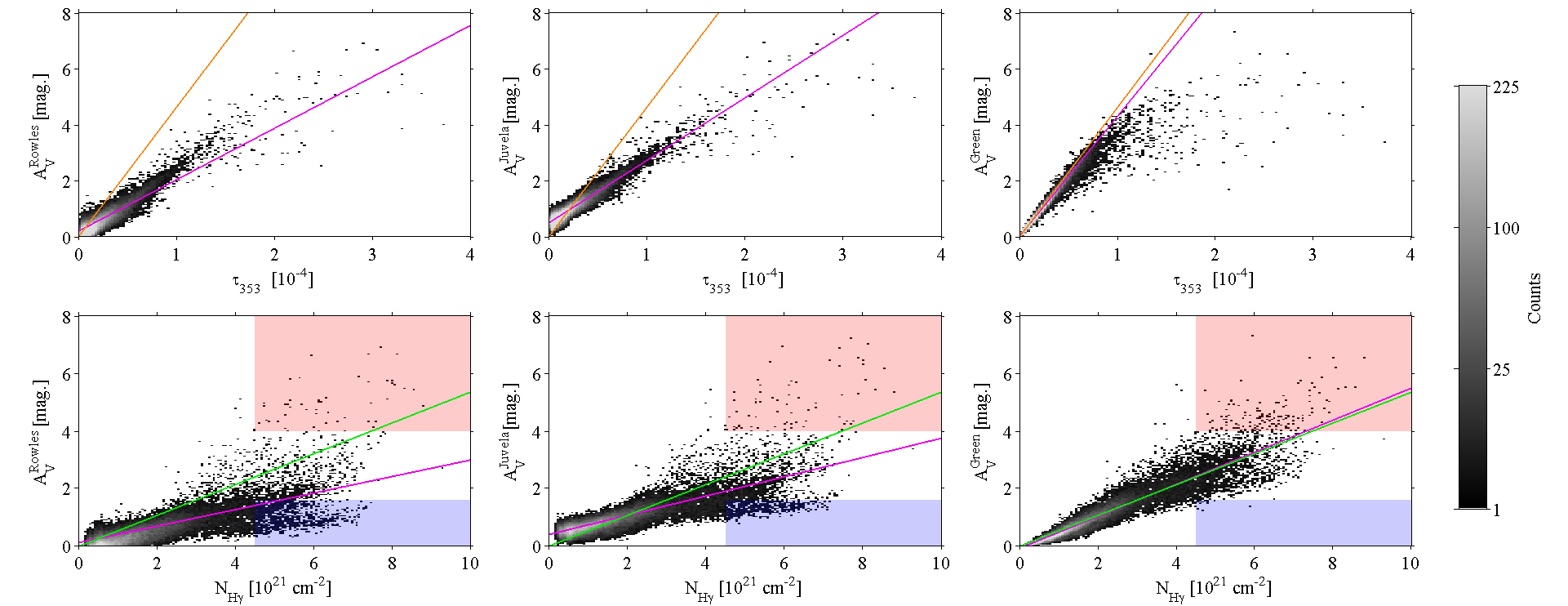}
  \caption{ \textit{Top}: 2D histograms of the correlations between the \av data and the dust optical depth at 353 GHz \citep{2014A&A...571A..11P} (\textit{Top}) or the total gas column density, \nhgam, measured with 0.4-100 GeV interstellar \g rays \citep{2017A&A...601A..78R} (\textit{Bottom}). All the maps are compared at a common angular resolution in 0\fdg25${\times}$0\fdg25 bins. The orange line in the top plots corresponds to a ratio \avtau$=4.62\times 10^4$ mag \citep{2014A&A...571A..11P}. The  green line in the bottom plots corresponds to a ratio \anh$=5.34\times $\ebvnhunit \citep{1978ApJ...224..132B}. In each panel the magenta line corresponds to the best fit of a linear regression. The blue and red areas correspond to the pixel selections used to generate the contours in Fig. \ref{fig:AVmaps}.}  
  \label{fig:AVnhtau}
\end{figure*}

We have selected the three following extinction maps :
\begin{itemize}
\item a visual extinction (\av) map constructed from the \textit{2MASS} stellar data, with the median colour excess of the 49 stars nearest to each direction \citep{2009MNRAS.395.1640R};
\item an extinction map in the J band, $A_{\rm J}$, derived from the \textit{2MASS} data with the NICER M2a processing at 12\arcmin-resolution \citep{2016A&A...585A..38J} ;
\item a three-dimensional \ebv map based on \mbox{\textit{Pan-STARRS} 1} and \textit{2MASS} stellar data integrated up to its maximal distance \citep{2015ApJ...810...25G}.
\end{itemize}
According to the extinction law of \cite{1989ApJ...345..245C}, we have converted the map of \cite{2016A&A...585A..38J} from $A_{\rm J}$ to \av with a multiplicative factor of 3.55 and we have applied the extinction factor $R_{\rm V}$=3.1 to convert the map of \cite{2015ApJ...810...25G} from reddening to visual extinction. The resulting three \av maps are presented in  \mbox{Fig. \ref{fig:AVmaps}} together with contours delimiting a selection of regions where \nh and \av are not linearly correlated. The pixel selections in \nh and \av that have been used to generate these contours are illustrated in Fig. \ref{fig:AVnhtau} as coloured areas. 
Figure \ref{fig:AVcomp} shows the 2D histograms of the correlations between each of the \av maps.
Figure \ref{fig:AVnhtau} shows the 2D histograms of the correlations between the \av maps and the dust optical depth at 353 GHz. It also displays the correlations with the gas column density derived from \g rays of interstellar origin in the 0.4-5 GeV energy band from \cite{2017A&A...601A..78R}. The datasets have been convolved to a common angular resolution and sampled in 0\fdg25${\times}$0\fdg25 bins in order to construct the histograms.

Figure \ref{fig:AVcomp} shows a good correlation between the visual extinctions obtained by \cite{2009MNRAS.395.1640R} and \cite{2016A&A...585A..38J}. The slope is close to one, but there is an offset of 0.5 mag in the latter. As the two maps are based on the same dataset, we suspect a method-dependent bias. We note the same offset in the visual extinctions of \cite{2016A&A...585A..38J} when compared to the map of \cite{2015ApJ...810...25G}. 

The map of \cite{2015ApJ...810...25G} presents less noise at low \av, but most of the pixels with \av$\gtrsim1$ mag exhibit extinctions above the values of the two others. Those differences may be explained by the use of optical data from \textit{Pan-STARRS} to complement the near-infrared data from \textit{2MASS}, especially in diffuse regions. The precision is improved by larger stellar statistics. Conversely, the near-infrared data from \textit{2MASS} is essential to scan deeper into dense molecular clouds. In the nearby clouds of Orion, Taurus and Perseus, \cite{2015ApJ...810...25G} obtain larger extinctions when combining \textit{2MASS} and \textit{Pan-STARRS} photometry than with the sole use of \textit{Pan-STARRS} \citep{2014ApJ...789...15S}. We also find a tighter correlation between the gas column densities derived in \g rays and the extinctions based on \textit{2MASS} and \textit{Pan-STARRS} data than with \textit{Pan-STARRS}-based extinctions (not illustrated here). 

Figure \ref{fig:AVnhtau} clearly shows that the dust optical depth strongly deviates from the visual extinction above 2 or 3 mag, independent of the method employed to derive the visual extinctions. At \taunu$>10^{-4}$ the optical depth non-linearly increases above the extinction expectation. At \av<2 mag, the distribution of \cite{2015ApJ...810...25G} compares well with the \avtau ratio of $4.62\times 10^4$ mag found by \cite{2014A&A...571A..11P} using colour-excess measurements toward 53 399 quasars and converted from \ebv to \av with $R_{\rm V}=3.1$. 

We note a significant saturation of the extinction at \nh$>4\times 10^{21}$ cm$^{-2}$ and \av$\approx1.5$ mag in the maps of \cite{2009MNRAS.395.1640R} and \cite{2016A&A...585A..38J}. This corresponds to regions in the Galactic disc (enclosed by blue contours in Fig. \ref{fig:AVmaps}). As noted by \cite{2016A&A...585A..38J}, near the Galactic plane the zero point of the extinction scale is not well defined because of the unknown mixture of stars and dust along the line of sight. This ambiguity could be resolved only with knowledge of the 3D distribution of the stars and the ISM. Indeed, the reddening map constructed in three dimensions by \cite{2015ApJ...810...25G} shows less deviation of the \anh ratio toward the Galactic plane.

At low gas column densities, \nh$<2\times 10^{21}$ cm$^{-2}$, the data from \cite{2016A&A...585A..38J} and \cite{2015ApJ...810...25G} show a better correlation with the gas column densities derived from the \g-ray emission. The slope is consistent with the standard ratio \anh$=5.34\times $\ebvnhunit of \cite{1978ApJ...224..132B}. At larger column densities, \nh$>4\times 10^{21}$ cm$^{-2}$, the dispersion in \av increases. Large extinctions deviating from the \nh trend are found in the densest cores of molecular clouds (within red contours in Fig. \ref{fig:AVmaps}). They are possibly due to the lack of suitable stars in heavily obscured regions.  


As the reddening map of \cite{2015ApJ...810...25G} presents a better correlation with both the dust optical depth and gas column densities, we have selected this map to study the relation between \nh and \ebv.
The extinction maps from \cite{2009MNRAS.395.1640R}, \cite{2016A&A...585A..38J}, and more generally all the 2D extinction maps derived from near-infrared surveys such as 2MASS are more suitable to study relative variations in extinction within a cloud or within a region of smaller angular size than the one studied here. Regions near the Galactic plane and very diffuse regions with extinctions lower than about 0.5 magnitude are more prone to extinction biases. 
3D methods adjusting simultaneously the distance and extinction of clouds along the lines of sight appear to be more robust than 2D methods at low Galactic latitudes \citep{2015ApJ...810...25G}. Additionally, the combination of optical and near-infrared surveys allows scanning a wider range in \av. On the one hand, optical surveys increase the stellar statistics in translucent regions, which help to reduce the noise level below 0.2 mag in \av. On the other hand, near-infrared surveys can probe deeper into the clouds and retain enough precision to \av $\sim 5$~mag.

\section{Best-fit coefficients of the $\gamma$-ray and dust models} 
\label{sec:annex}

\renewcommand{\arraystretch}{1.7}
\setlength{\tabcolsep}{0.27cm}
\begin{table*}
\caption{Best-fit coefficients of the $\gamma$-ray model in each energy band and of the dust model. }
\centering
\begin{tabular}{l|r r r r r|| l | r}
\hline \hline
$\gamma$-ray & \multicolumn{5}{c||}{Energy band [MeV]}& \multicolumn{2}{c}{$E$(B$-$V)}\\
model & $10^{2.6}-10^{2.8}$& $10^{2.8}-10^{3.2}$&$10^{3.2}-10^{3.6}$&$10^{3.6}-10^5$&$10^{2.6}-10^5$ & \multicolumn{2}{c}{model}  \\
\hline
$q_{\rm{HI \; Cet}}   $ & 1.17$\pm$0.06 & 1.10$\pm$0.05 & 1.11$\pm$0.05 & 1.02$\pm$0.12 & 1.08$\pm$0.03&$y_{\rm{HI \; Cet}}^{\rm a}$ & 1.61$\pm$0.01 \\ 
$q_{\rm{HI \; TauS}}   $ & 1.22$\pm$0.02 & 1.23$\pm$0.02 & 1.17$\pm$0.02 & 1.16$\pm$0.04 & 1.20$\pm$0.01&$y_{\rm{HI \; TauS}}^{\rm a}$ & 1.96$\pm$0.01 \\ 
$q_{\rm{HI \; TauN}}   $ & 1.10$\pm$0.02 & 1.04$\pm$0.02 & 1.06$\pm$0.02 & 1.13$\pm$0.05 & 1.05$\pm$0.01&$y_{\rm{HI \; TauN}}^{\rm a}$ & 1.70$\pm$0.01 \\ 
$q_{\rm{HI \; TauM}}   $ & 1.26$\pm$0.02 & 1.23$\pm$0.02 & 1.19$\pm$0.02 & 1.23$\pm$0.04 & 1.23$\pm$0.01&$y_{\rm{HI \; TauM}}^{\rm a}$ & 1.89$\pm$0.03 \\ 
$q_{\rm{HI \; Cal}}   $ & 1.19$\pm$0.05 & 1.18$\pm$0.03 & 1.08$\pm$0.04 & 1.15$\pm$0.08 & 1.16$\pm$0.02&$y_{\rm{HI \; Cal}}^{\rm a}$ & 2.11$\pm$0.01 \\ 
$q_{\rm{HI \; Per}}   $ & 1.58$\pm$0.14 & 1.32$\pm$0.09 & 1.39$\pm$0.11 & 1.08$\pm$0.20 & 1.34$\pm$0.06&$y_{\rm{HI \; Per}}^{\rm a}$ & 2.55$\pm$0.10 \\ 
$q_{\rm{HI \; Gal}}   $ & 0.57$\pm$0.02 & 0.60$\pm$0.02 & 0.57$\pm$0.02 & 0.65$\pm$0.04 & 0.57$\pm$0.01&$y_{\rm{HI \; Gal}}^{\rm a}$ & 0.68$\pm$0.06 \\ 
$q_{\rm{CO \; Cet}}$ & 1.32$\pm$0.71 & 2.51$\pm$0.47 & 2.43$\pm$0.51 & 0.00$\pm$0.89 & 2.16$\pm$0.30&$y_{\rm{CO \; Cet}}^{\rm b}$ & 4.64$\pm$0.03 \\ 
$q_{\rm{CO \; TauS}}$ & 1.91$\pm$0.20 & 2.10$\pm$0.15 & 2.08$\pm$0.16 & 1.68$\pm$0.30 & 2.01$\pm$0.09&$y_{\rm{CO \; TauS}}^{\rm b}$ & 4.14$\pm$0.02 \\ 
$q_{\rm{CO \; TauN}}      $ & 1.62$\pm$0.10 & 1.62$\pm$0.07 & 1.59$\pm$0.08 & 1.67$\pm$0.16 & 1.63$\pm$0.05&$y_{\rm{CO \; TauN}}^{\rm b}$ & 4.16$\pm$0.03 \\ 
$q_{\rm{CO \; TauM}}   $ & 1.81$\pm$0.05 & 1.72$\pm$0.04 & 1.64$\pm$0.05 & 1.69$\pm$0.09 & 1.68$\pm$0.03&$y_{\rm{CO \; TauM}}^{\rm b}$ & 3.67$\pm$0.03 \\ 
$q_{\rm{CO \; Cal}}   $ & 2.00$\pm$0.10 & 1.92$\pm$0.07 & 1.81$\pm$0.08 & 1.81$\pm$0.16 & 1.85$\pm$0.05&$y_{\rm{CO \; Cal}}^{\rm b}$ & 3.92$\pm$0.23 \\ 
$q_{\rm{CO \; Per}}   $ & 1.48$\pm$0.12 & 1.63$\pm$0.08 & 1.70$\pm$0.09 & 1.97$\pm$0.17 & 1.64$\pm$0.05&$y_{\rm{CO \; Per}}^{\rm b}$ & 2.52$\pm$0.14 \\ 
$q_{\rm{CO \; Gal}}$ & 1.98$\pm$0.59 & 1.49$\pm$0.39 & 2.09$\pm$0.42 & 1.61$\pm$0.74 & 1.58$\pm$0.25&$y_{\rm{CO \; Gal}}^{\rm b}$ & 7.44$\pm$0.02 \\ 
$q_{\rm{ff} }     $ & 2.34$\pm$0.69 & 3.14$\pm$0.51 & 1.58$\pm$0.55 & 3.27$\pm$1.17 & 2.56$\pm$0.32&$y_{\rm{ff} }^{\rm c}     $ & 2.09$\pm$0.01 \\ 
$q_{\rm{COsat}}   $ & 55.27$\pm$3.40 & 60.38$\pm$2.42 & 56.62$\pm$2.66 & 61.48$\pm$4.72 & 59.67$\pm$1.53&$y_{\rm{COsat}}^{\rm a}   $ & 2.92$\pm$0.01 \\ 
$q_{\rm{DNM}}      $ & 45.82$\pm$1.46 & 44.53$\pm$1.06 & 43.23$\pm$1.19 & 46.84$\pm$2.38 & 45.03$\pm$0.67&$y_{\rm{DNM}}^{\rm a}$ & 2.77$\pm$0.01 \\ 
$q_{\rm{ic}}      $ & 2.88$\pm$0.77 & 4.54$\pm$0.64 & 3.88$\pm$0.65 & 2.48$\pm$0.84 & 4.20$\pm$0.36&$y_{\rm{iso}}^{\rm d}$ & -2.61$\pm$0.01 \\ 
$q_{\rm{iso}}      $ & 1.39$\pm$0.25 & 1.12$\pm$0.28 & 0.70$\pm$0.28 & 1.70$\pm$0.27 & 1.07$\pm$0.14 \\

\hline \hline
\end{tabular}
\vspace{0.3cm}
\tablefoot{The q coefficients are expressed in $10^{20}$ cm$^{-2}$ (K km s$^{-1}$) for the CO; 3.8 10$^{15}$ cm$^{-2}$Jy$^{-1}$sr for the free-free; $10^{20}$ cm$^{-2}$ mag.$^{-1}$ for the DNM and CO$_{Sat}$; other q coefficients are simple normalization factors. \\
The y coefficients are expressed as follows: $^{{\rm a}}$ in $10^{-22}$ cm$^{2}$;
$^{{\rm b}}$ in $10^{-2}$ mag. K$^{-1}$km$^{-1}$s;
$^{{\rm c}}$ in 3.8 $10^{-7}$ mag. Jy$^{-1}$sr;
$^{{\rm d}}$ in $10^{-2}$ mag. \\
The uncertainties reported are statistical only; the systematic uncertainties are ~5\% (based on the effective area uncertainties)\footnote{\url{https://fermi.gsfc.nasa.gov/ssc/data/analysis/LAT_caveats.html}}. }

 \label{tab:fitscoef}
 \end{table*}

\end{document}